\documentclass[preprint,11pt]{elsarticle}

\usepackage{hyperref}
\usepackage{lineno}
\usepackage{multirow}
\usepackage{multicol}
\usepackage{gensymb}
\usepackage{amssymb}
\usepackage{tablefootnote}
\usepackage{caption}
\usepackage{subcaption}
\usepackage{wrapfig}
\usepackage{amsmath}
\usepackage[dvipsnames]{xcolor}
\usepackage{comment}

\usepackage{dblfloatfix}
\usepackage{fixltx2e}
\usepackage{placeins}

\newcommand{\micronsqs}{$\mu$m$^2$~}
\newcommand{\micronsq}{$\mu$m$^2$}
\newcommand{\microns}{$\mu$m~}
\newcommand{\micron}{$\mu$m}
\newcommand{\sigmahits}{$\sigma_{hit\; pos}$~}
\newcommand{\sigmamss}{$\sigma_{MS}$~}

\begin{document}

\begin{frontmatter}

\title{High-Precision 4D Tracking with Large Pixels using Thin Resistive Silicon Detectors}

\author[1,2]{R.~Arcidiacono\corref{corrauth}}
\cortext[corrauth]{Corresponding author}
\ead{arcidiacono@to.infn.it}
\author[4]{G. Borghi}
\author[address2]{M. Boscardin}
\author[1]{N.Cartiglia}
\author[address2]{M. Centis Vignali}
\author[3]{M. Costa}
\author[address5]{G-F. Dalla Betta}
\author[1]{M.~Ferrero}
\author[address2]{F. Ficorella}
\author[3]{G.~Gioachin}
\author[3]{L.~Lanteri}
\author[1]{M.~Mandurrino}
\author[1]{L.~Menzio}
\author[1,3]{R.~Mulargia}
\author[address5]{L. Pancheri}
\author[address2]{G. Paternoster}
\author[1]{A. Rojas}
\author[address6]{H-F W. Sadrozinski}
\author[address6]{A. Seiden}
\author[1]{F.~Siviero}
\author[1,3]{V.~Sola}
\author[1,3]{M.~Tornago}

\address[1]{INFN, Sezione di Torino, Italy}
\address[2]{Universit\`a del Piemonte Orientale, Italy}
\address[3]{Universit\`a di Torino, Torino, Italy}
\address[4]{Politecnico di Milano, Milano, Italy}
\address[address2]{Fondazione Bruno Kessler, Trento, Italy}
\address[address5]{Universit\`a degli Studi di Trento, Trento, Italy}
\address[address6]{University of California at Santa Cruz, CA, US}

\begin{abstract}
The basic principle of operation of silicon sensors with resistive read-out is built-in charge sharing. Resistive Silicon Detectors (RSD, also known as AC-LGAD), exploiting the signals seen on the electrodes surrounding the impact point, achieve excellent space and time resolutions even with very large pixels. In this paper, a TCT system using a 1064 nm picosecond laser is used to characterize sensors from the second RSD production at the Fondazione Bruno Kessler.
The paper first introduces the parametrization of the errors in the determination of the position and time coordinates in RSD, then outlines the reconstruction method, and finally presents the results.  Three different 
pixel sizes are used in the analysis: 200 $\times$ 340,  450 $\times$ 450, and 1300 $\times$ 1300 \micronsq. At gain = 30, the 450 $\times$ 450 \micronsqs pixel achieves a time jitter of 20 ps and a spatial resolution of 15 \microns concurrently, while the  1300 $\times$ 1300 \micronsqs pixel  achieves 30 ps and 30 \micron, respectively. The implementation of cross-shaped electrodes improves considerably the response uniformity over the pixel surface.

\end{abstract}


\end{frontmatter}



\section{Introduction}
High-precision tracking requires the concurrent minimization of two quantities:  (i) the hit position resolution \sigmahits,  how precisely the impact point is located on the sensor surface, and (ii) the multiple scattering position resolution \sigmamss, how much the tracker materials (cables, cooling,
mechanics, the detector itself) influence the determination of the hit position.

\begin{figure}[htb]
\begin{center}
\includegraphics[width=0.95\textwidth]{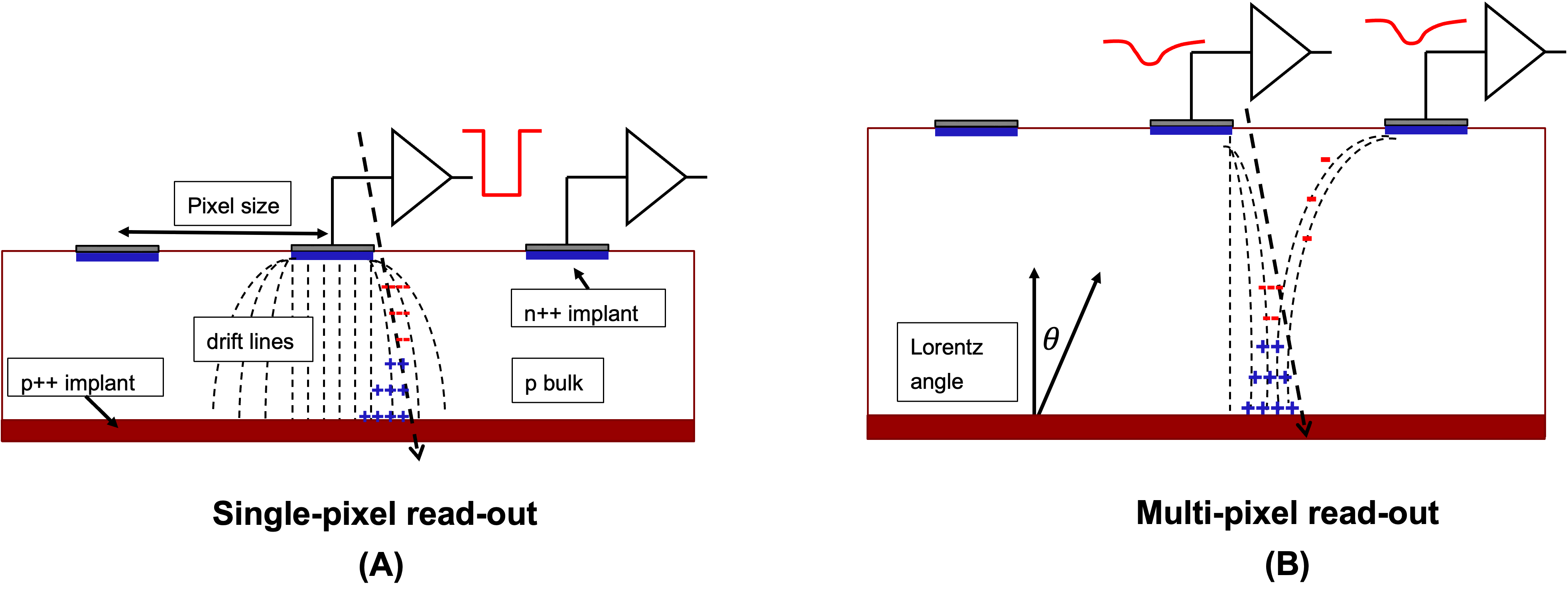}
\caption{(A) Single and (B) multi-pixel read-out schemes for silicon sensors. The presence of a magnetic field modifies the drift line adding a Lorentz angle that induces charge sharing between two adjacent pads.}
\label{fig:det}
\end{center}
\end{figure}

 The two terms,  \sigmahits and \sigmamss, are deeply linked to each other and to the type of read-out architecture (single or multi-pixels) used in the system.
In single-pixel read-out, shown in Figure~\ref{fig:det}(A), the hit resolution \sigmahits  is the standard deviation of a uniform random variable distributed over the pixel size, \sigmahits = $k*pixel\;size/\sqrt{12}$,   where $k  \sim 0.5 - 1$. This relationship is at the root of the limited spatial accuracy achievable with single-pixel read-out: the pixel size determines the spatial resolution. Only tiny pixels (25 $\times$ 25 \micronsq) achieve a precision of  5-10 \micron, and it is practically impossible to reach better resolutions.

In multi-pixels read-out, shown in Figure~\ref{fig:det}(B), the signal is split between two (even three) pixels, and the position of a hit can be calculated as the signal-weighted centroid (or a similar algorithm)  of the two pixels coordinates. This method is robust and reaches excellent accuracy, yielding \sigmahits significantly smaller than $k*pixel \; size/\sqrt(12)$ . However, sharing requires large signals and, therefore, thick sensors to maintain full detection efficiency even when the signal is split. When signal sharing is obtained via the introduction of a magnetic field, Figure~\ref{fig:det} (B), the sensor needs to be even thicker (200-300 \micron) to allow sufficient bending of the drift lines. As two examples of these approaches, the ATLAS experiment uses a vertex tracker with small pixels and a single-pixel read-out (50 $\times$ 50 \micronsqs pixels and \sigmahits = 5 \microns) \cite{Collaboration_2008} while the CMS experiment has chosen larger pixel and a multi-pixels read-out with thick sensors and a strong magnetic field (100 $\times$ 150 \micronsqs pixels and \sigmahits = 5 \micron, 3.8 Tesla magnet) to exploit charge sharing in the determination of the hit position \cite{Agram:2012qba}.

Interestingly,  the very mechanism that optimizes \sigmahits is detrimental to \sigmamss: thick sensors, necessary for signal sharing, cause significant multiple scattering and deteriorate the overall accuracy of the tracker system. 
In this paper, the performance of Resistive Silicon Detectors (RSD) are presented, and the results demonstrate that this novel design minimizes at the same time \sigmahits and \sigmamss,  while using large pixels, a key feature to reduce power consumption. 

\section{RSD principles of operation}

\begin{figure}[htbp]
\begin{center}
\includegraphics[width=0.95\textwidth]{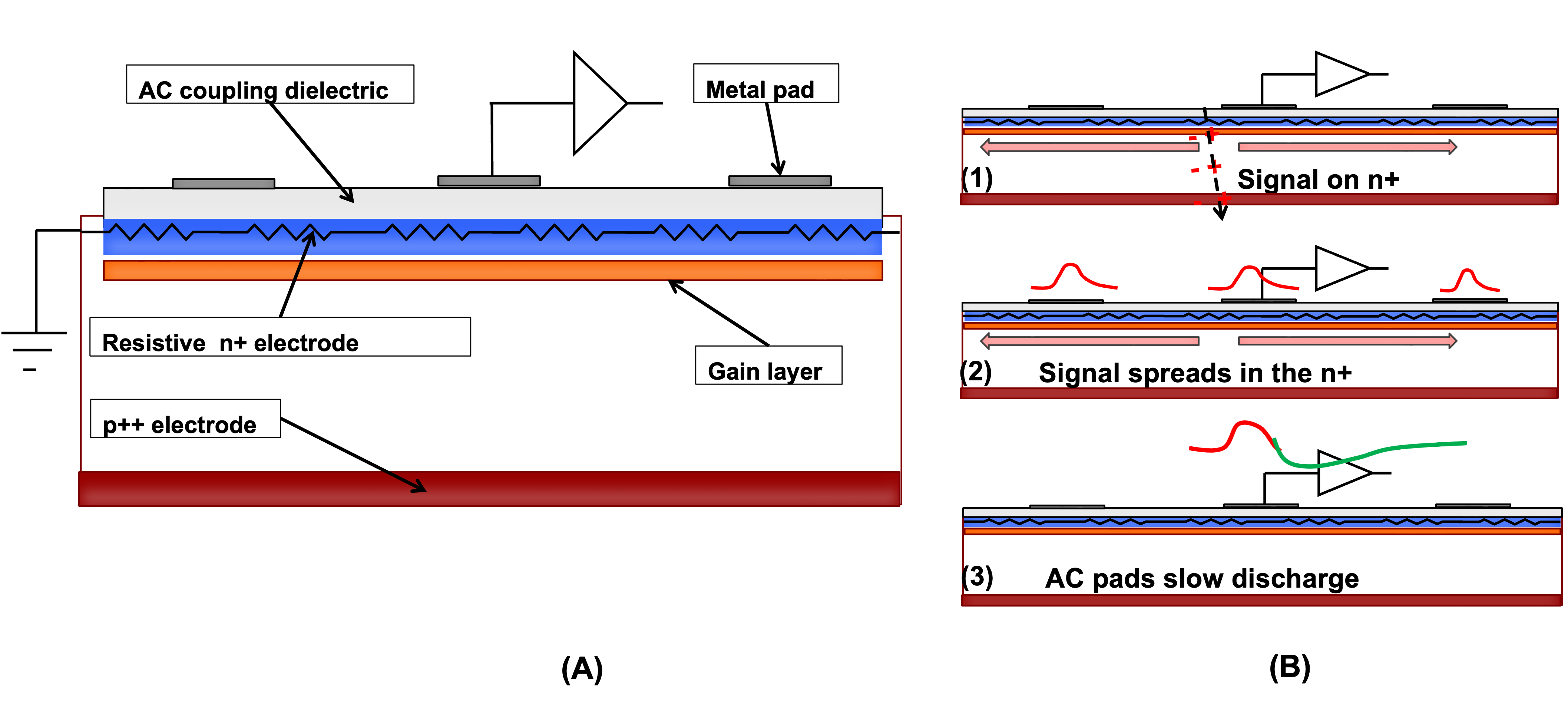}
\caption{(A) Sketch of an RSD. The main components are indicated in the sketch. (B) The three phases of signal formation in RSDs. }
\label{fig:rsd}
\end{center}
\end{figure}

RSDs are thin silicon sensors that combine two design innovations~\cite{tornago2020resistive}:  (i) built-in signal sharing due to the presence of resistive read-out and (ii)  internal gain due to the adoption of the low-gain avalanche diode design. Figure~\ref{fig:rsd}(A) shows a sketch of the RSD design, while Figure~\ref{fig:rsd}(B) outlines the working principles: (1) the drift of the e/h pairs generates an induced signal on the n$^+$ resistive layer, the signal is boosted by the presence of an internal gain mechanism. (2) The signal spreads toward the ground in the n$^+$ resistive layer; the fast component of the signal is visible on the AC metal pads as they offer the lowest impedance high-frequency paths to ground. (3) The AC pads discharge with a time
constant that depends on the read-out input resistance, the n$^+$ sheet resistance, and the system capacitance. 

\begin{figure}[htb]
\begin{center}
\includegraphics[width=0.95\textwidth]{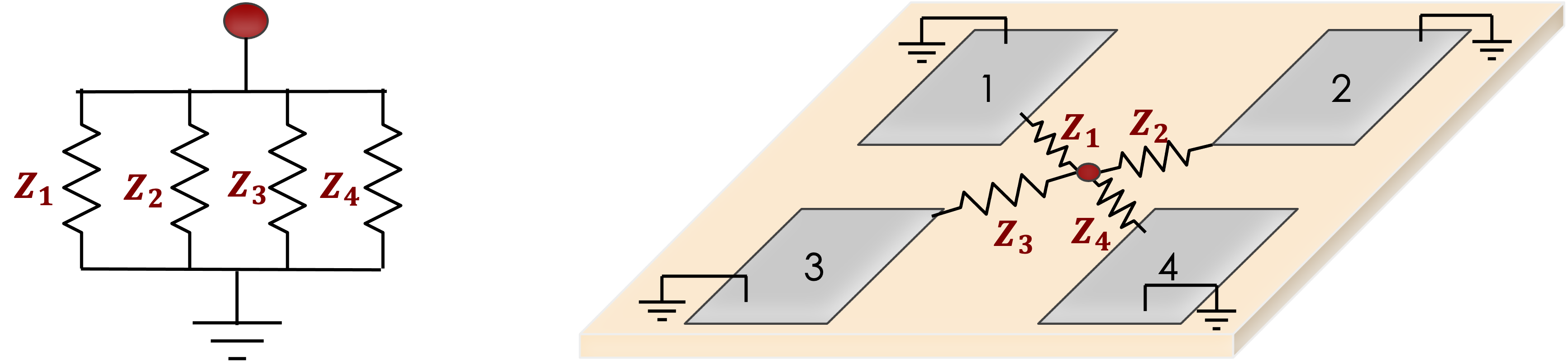}
\caption{In RSD, the signal splits among the read-out pads as a current in an impedance divider}
\label{fig:split}
\end{center}
\end{figure}

The signal splits among the read-out pads as a current in an impedance divider, where the impedance is that of the paths connecting the impact point to each of the read-out pads, as sketched in Figure~\ref{fig:split}.

\section{Parametrisation of the spatial resolution of RSD}

\begin{center}
\includegraphics[width=0.9\textwidth]{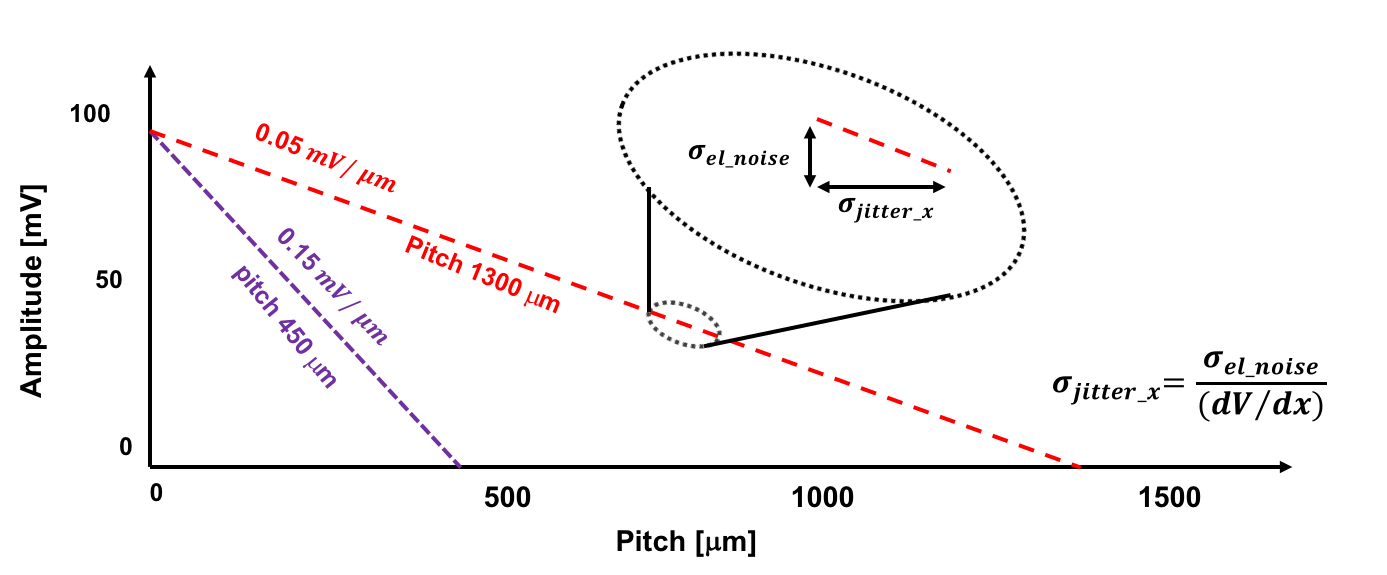}
\captionof{figure}{Amplitude as a function of distance for two RSD geometries. In large structures, the decrease of the signal per micron is smaller, leading to a larger the jitter term. }
\label{fig:resol}
\end{center}

There are four distinct contributions to the spatial resolution, listed in Equation~\ref{eq:spaceres}:

\begin{equation}
\label{eq:spaceres}
\sigma_{hit\; pos}^2 =  \sigma_{jitter}^2 +  \sigma_{rec}^2 +  \sigma_{setup}^2 +\sigma_{sensor}^2.
\end{equation}
The first contribution, $ \sigma_{jitter}$, degrades the precision of the measurement, while the other terms degrade the accuracy.

\begin{itemize}
\item $\sigma_{jitter}$: this term is related to the electronic noise $\sigma_{el-noise}$. As illustrated in  Figure~\ref{fig:resol}, the variation of signal amplitude  due to the electronic noise induces uncertainty in the hit localization given by 
\begin{equation}
\label{eq:jitterxx}
 \sigma_{jitter} = \sigma_{el-noise}/(dV/dx) \sim  \frac{\sigma_{el-noise}}{Amplitude}\times pixel  \;size,
 \end{equation}
 where $dV/dx$ depends upon the signal amplitude and pixel size. In the examples shown in Figure ~\ref{fig:resol},  the amplitude changes by dV/dx =  0.15 (0.05)  mV/\microns for a  450 (1300) \microns pixel: assuming $\sigma_{el-noise}\sim$ 2 mV, the jitter term is about $\sigma_{jitter} \sim$ 13 \microns for the 450 \microns pixel, while it becomes $\sigma_{jitter} \sim$ 40 \microns for the  1300 \microns structure. When the signal is split among $n$ read-out channels, the amplitude seen on each pad  is actually 1/$n$ smaller, while the effective noise is reduced by $\sqrt{n}$ due to the combination of  the signals from the $n$ read-out pads:
\begin{equation}
\label{eq:jitterx}
\sigma_{jitter}\sim \frac{  \frac{\sigma_{el-noise}}{\sqrt{n}} } { \frac{dV/dx}{n}} = \frac{\sigma_{el-noise}}{dV/dx}\sqrt{n}. 
\end{equation}

As seen in this paragraph, the electronic noise sets the limit of the spatial precision, and for equal noise, the precision depends linearly on the pixel size. If high spatial precision with large pixels is needed, then the electronics should be very low noise and the signal gain large enough.
\item $  \sigma_{rec}$: the reconstruction code uses algorithms to infer the hit position from the measured signals. This can be done in several ways: analytically, using methods based on look-up tables, or with more advanced techniques such as machine learning. In all methods, the reconstructed hit positions might have a position-dependent systematic offset with respect to the true position. 
\item $  \sigma_{setup}$: this term includes the uncertainties related to the experimental set-up. Specifically, the most important are those effects that change the relative amplitude between the actual signal sharing the measured signal sharing (for example, differences in the amplifier gain used to read out the electrodes).

\item $\sigma_{sensor}$ this term groups all sensor imperfections contributing to an uneven signal sharing among pads. The most obvious one is a varying n$^+$ resistivity: a 2\% difference in n$^+$ resistivity turns an equal signal split between two pads, 50 mV on each of the two pads, into a 49.5 mV - 50.5 mV split, yielding a shift of the position assignment of  $\sim$ 7 \microns for the 450 \microns geometry and   20 \microns for the 1300 \microns design. The uniformity of the n$^+$ resistive layer (and that of the gain implant) is a crucial parameter in RSD optimized for micron-level position resolution.

\end{itemize}

\section{Parametrisation of the time resolution of RSD}

The parametrization expressing the time resolution of a single read-out pad is similar to that of standard UFSD,  a complete explanation of the contributions can be found in~\cite{ROPP}. In RSD, there is an  additional contribution due to the uncertainty  in the determination of the signal delay, i.e. the time interval between the hit time and when the signal is visible on the read-out pad.

\begin{equation}
\label{eq:timeres}
\sigma_{hit\; time}^2 =   \sigma_{jitter}^2 + \sigma_{Landau}^2 + \sigma_{delay}^2
\end{equation}

\begin{itemize}
\item $\sigma_{jitter}$:  due to the electronic, $\sigma_{jitter} = \sigma_{el-noise}$/(dV/dt) 
\item $\sigma_{Landau}$: due to non-uniform ionization. Assuming a 50 \microns thick sensor, this term is about 30 ps.
\item $ \sigma_{delay}$:  due to the uncertainty on the hit position reconstruction. 
\end{itemize}
Overall, a good time resolution requires large signals, low noise electronics, thin sensors, and a good determination of the impact point. 

In RSDs, the time resolution is limited by the time jitter term for small signal, and  by the Landau noise for large signals, while it is not degraded significantly by moderate sensor non-uniformity or by an uncertainty in the hit position of the order of 30-40 \micron. If the $n^+$ resistivity is low, and the delay is well measured, the time resolution does not depend significantly  on the pixel size. As for the spatial case, see Equation~\ref{eq:jitterx}, $\sigma_{jitter}$  increases when splitting the signal on $n$ read-out pads as $\sigma_{jitter} \propto \sqrt{n}$.

\section{The second RSD production (RSD2) at Fondazione Bruno Kessler}

\begin{figure}[htb]
\begin{center}
\includegraphics[width=0.95\textwidth]{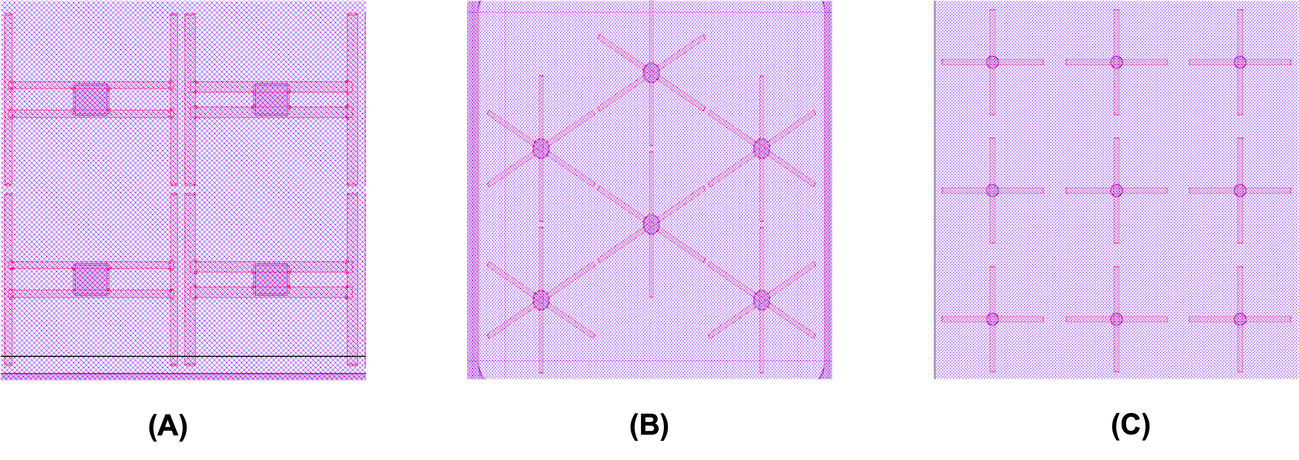}
\caption{ (A) 2-pixel sharing: this configuration is useful when only one coordinate needs to be determined accurately 
(B) 3-pixel sharing: the electrodes are at the vertexes of a triangle, with arms extending out. (C) 4-pixel sharing: the electrodes are placed at the vertexes of a square.}
\label{fig:shapes}
\end{center}
\end{figure}

The studies performed using the first production of resistive silicon detectors~\cite{8846722} (RSD1), manufactured at Fondazione Bruno Kessler (FBK), have shown that signal sharing can be optimized by a careful design of the read-out electrode shape~\cite{tornago2020resistive, Siviero_2021}. The electrodes need to surround as much as possible the pixel area to confine the signal spread to a pre-defined number of pads, and the metal of the electrodes needs to be minimized to achieve a uniform response over the pixel area. The RSD2 production includes several optimizations of the electrode shapes~\cite{Arcidiacono_2022}, a few examples are shown in  Figure~\ref{fig:shapes}. A two-electrodes configuration (A) is particularly suited when only one of the two coordinates needs to be known precisely, for example, in the measurements of the trajectory of a particle in a magnetic field.   Configurations (B) and (C) split the signal respectively among three or four electrodes, with (C) sharing it more uniformly due to a larger angle between the electrode arms. For each of these configurations, several design variations have been implemented in RSD2, changing the arm width, the distance between arms, and the size of the contact pad. These aspects impact the electrode capacitance, the shape of the signal, and the capability of limiting the signal spread outside the pixel area. 

The analysis presented in this paper uses three different versions of the type shown in Figure~\ref{fig:shapes} (C), listed in Table~\ref{tab:struc}.

\begin{table*}[htb]
  \centering
  \tabcolsep3pt
  \begin{tabular}{|c|c|c|c|c|c|}
  \hline
    Name & sensor  & pixel  & contact pad   & arm width  & gap between arms \\  
        & 	[\micronsq]  & [\micronsq]  &   [\micron]  & [\micron] & [\micron] \\ \hline\hline   
    A & 800 $\times$ 800 & 200 $\times$ 345  & 30 & 10 & 5 \\	\hline
       B & 2700 $\times$ 2700 & 450 $\times$ 450  & 45 & 20 & 10 \\	\hline	
             C &  2700 $\times$ 2700& 1300 $\times$ 1300 & 90 & 20 & 100 \\	\hline		
       \end{tabular}
  \caption{List of the parameters of the structures used in this analysis. The three structures use the cross-shaped electrodes shown in Figure~\ref{fig:shapes}(C).}
  \label{tab:struc}
\end{table*}

\section{The experimental set-up}
\label{sec:exset-up}
The present studies have been performed with a high-precision  Transient Current Technique (TCT) set-up \cite{particulars_1}. In this set-up, a   pico-laser with a 1064 nm wavelength generates e/h pairs in the sensor under test, emulating the passage of a minimum ionizing particle. The diameter of the laser spot has been measured to be in the 5 -10 \microns range, depending on the precision of the calibration procedure. The sensors were tested using a 16-channel read-out board designed at Fermilab (the so-called FNAL board). Each read-out channel consists of a 2-stage amplifier chain based on the Mini-Circuits GALI-66+ integrated circuit with a 25 $\Omega$ input resistance, a $\sim 5 \; k\Omega$ total trans-resistance, and a bandwidth of $1 \; GHz$ \cite{apresyan2020measurements}.  The amplified signals were then recorded for offline analyses by a fast digitizer (16-channel CAEN DT5742,  with a 5 GS/s sampling rate). The noise of the system, as measured using empty events, was evaluated to be $\sigma_{el-noise}$ = 1.04 mV.  In the position and time reconstruction of the events, only the signals collected on four read-out pads at the corner of the pixel under study were used.

 A key point when using the TCT set-up is the calibration of the laser intensity. Since the performance of the FNAL board are very well measured,  the laser intensity has been monitored by measuring the area of the signal generated on the $n^{+}$ resistive layer, the so-called DC-electrode, knowing that a signal area of 50 picoWeber corresponds to about 1 fC of charge. Using this calibration and the gain-bias characteristics of the sensor under study, it is straightforward to set the laser intensity so that it generates a 1-MIP equivalent charge.

\section{The reconstruction method}
\label{sec:rec}

\begin{figure}[htb]
\begin{center}
\includegraphics[width=0.95\textwidth]{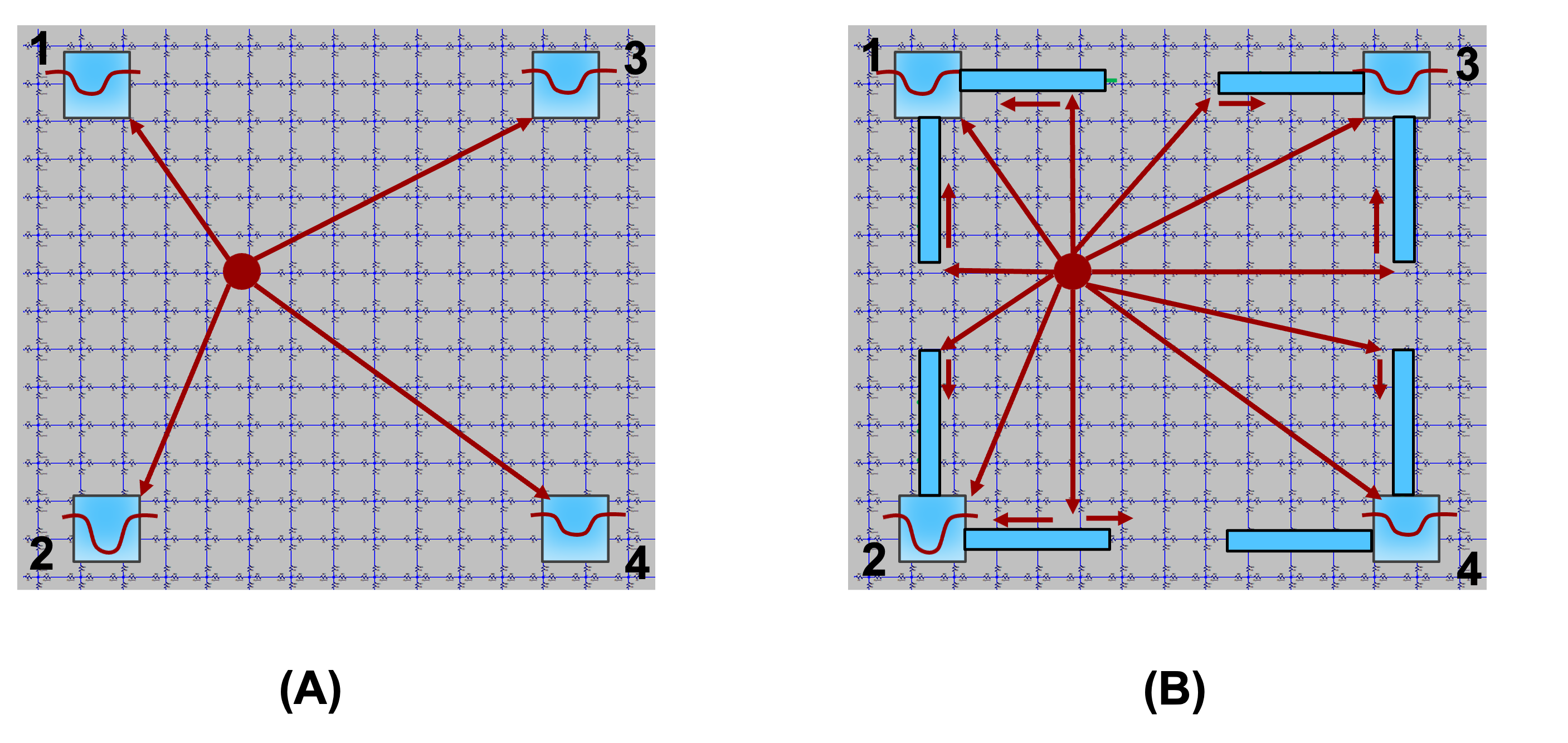}
\caption{Sketch of signal sharing for two RSD with different layouts of the read-out electrodes. (A) a layout with point-like electrodes. (B) a layout with extended electrodes. An analytic formulation of the sharing mechanism for (B) is difficult to be achieved due to the presence of multiple current paths leading to the same read-out electrode.   }
\label{fig:disp}
\end{center}
\end{figure}

In RSDs, the way the signal is shared among the read-out electrodes depends upon the relative distances between the impact point and the read-out electrodes. 
For this reason,  the position of the impact point can be identified using the measured signal sharing. For specific electrode layouts, such as the one shown in Figure~\ref{fig:disp}(A), the distance between the hit position and each of the pads is uniquely identified. In this configuration, it is possible to calculate the signal split and delays, as performed in \cite{tornago2020resistive}, and infer the position of the impact point by comparing the measured and calculated signal sharing. 
On the other hand, for layouts with extended electrodes, like the one shown in Figure~\ref{fig:disp}(B), the analytic approach does not model well enough the propagation on the resistive layer. The signal on a given pad is, in this case, the sum of many contributions, each following a different path. For such layouts, an efficient approach is to identify an appropriate reconstruction algorithm and then correct its biases by measuring them experimentally.

\subsection{Reconstruction of the hit position}
The first step in the position determination for the case of Figure~\ref{fig:disp}(B) is to define the reconstruction algorithms. For this analysis, two different algorithms were considered: (i) the Signal-Weighted Position (SWP), and  (ii)  the Discretized Position Circuit (DPC)~\cite{507162}. 

The SWP  equations are:
\begin{equation}
\begin{aligned}
\label{eq:swp}
 x_{meas} = \frac{\sum_i^4 x_i*A_i} { \sum_i^4 A_i} \\
 y_{meas} = \frac{\sum_i^4 y_i*A_i} { \sum_i^4 A_i}, \\
\end{aligned}
\end{equation}
where $x_{meas}, y_{meas}$ are the hit coordinates, $x_i, y_i $  the coordinates of the pads central points, and  $A_i$ the signal measured on pad $i$.

 In DCP,  the position is reconstructed using the signal unbalance between  the two sides (right - left, top - bottom) of the pixel, as shown in Equation~\ref{eq:dpc}:
\begin{equation}
\begin{aligned}
\label{eq:dpc}
 x_{meas} = x_0 +  k_x*\frac{ (A_3 + A_4) - (A_1+A_2) }{ A_1+A_2+A_3+A_4} \\
 y_{meas} = y_0 +  k_y*\frac{ (A_1 + A_3) - (A_2+A_4) }{ A_1+A_2+A_3+A_4} ,
\end{aligned}
\end{equation}

where $A_i$ is the signal measured on the pad $i$, $x_0, y_0$ are the coordinates of the central point of the pixel, and $k_x, k_y$ are given by:

\begin{equation}
\begin{aligned}
\label{eq:kdpc}
k_x = \frac{pixel \; size}{2}* \frac{1}{\frac{ (A_3 + A_4) - (A_1+A_2) }{ A_1+A_2+A_3+A_4} |_{x =x_3}}\\
k_y = \frac{pixel \; size}{2}* \frac{1}{\frac{ (A_1 + A_3) - (A_2+A_4) }{ A_1+A_2+A_3+A_4} |_{y = y_3}}.
\end{aligned}
\end{equation}

The coefficients $k_x, k_y$ are measured experimentally and account for the fact that if the hit point is on one side of the pixel (at $x = x_3$ to determine $k_x$ and at $y = y_3$ to determine $k_y$), the signals measured on the read-out pads on the other side might not be equal to zero. This is especially important in small pixels: in that case, if $k_x, k_y$ are set to 1, the reconstruction algorithm clusters the hit positions toward the center of the pixel.  

The quantity $A_i$ in both SWP and DPC can be either the amplitude or the area of the signal. One important difference between the two quantities is that 
 the amplitude of a signal decreases during the propagation on the n$^+$ resistive layer while the area does not change. 
 
 Amplitudes,  therefore, carry more information than areas and potentially lead to a better resolution. Ultimately,  the decision to use areas or amplitudes depends on the type of electronics used, i.e., on the signal-to-noise ratios of the two choices.

\subsubsection{Accuracy of the reconstruction methods}
The next step is to evaluate the accuracy of the reconstruction methods (SWP and DPC), i.e., to measure by how much the measured coordinates ($x_{meas}, y_{meas}$) differ systematically from the true hit coordinates ($x_{true}, y_{true}$). This step is performed by collecting data, called in the following  "training data",  with the TCT set-up.

\begin{figure}[htb]
\begin{center}
\includegraphics[width=0.95\textwidth]{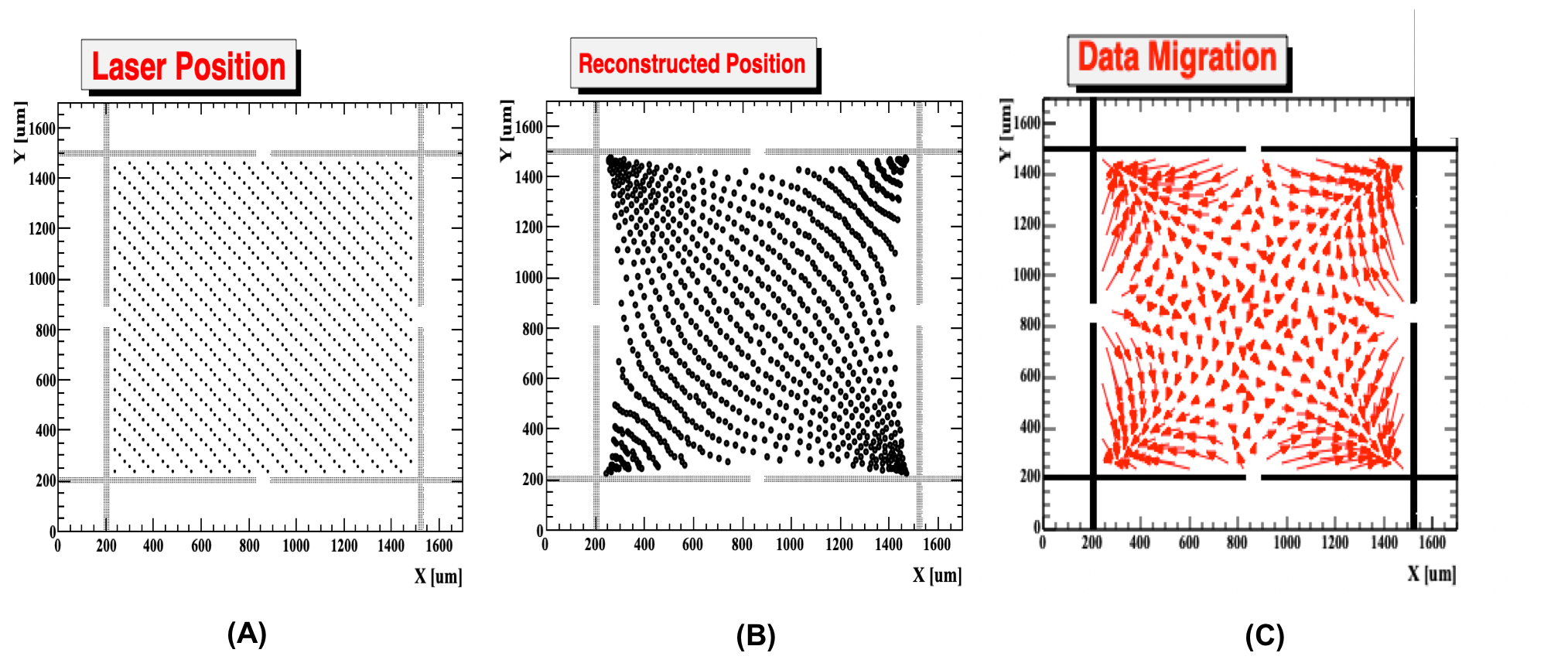}
\caption{Determination of the migration map (for DPC using amplitude)  for a 1300 $\times$ 1300 \micronsqs pixel:  (A) map of the laser positions on the pixel, (B) map of the reconstructed positions, (C) migration map obtained connecting the true positions with the reconstructed positions. }
\label{fig:data}
\end{center}
\end{figure}

In each acquisition sequence, the laser moves by 10 or 20 \microns covering the whole DUT surface, and for each position, 100 shots are recorded. The hit positions are then reconstructed either using SWP or DPC and compared with the true coordinates. Figure~\ref{fig:data} shows an example of this process: (A) map of the laser positions covering the surface of the pixel  (B) map of the reconstructed positions  (C) the migration map obtained connecting the true positions with the reconstructed positions:  it represents graphically the offset associated to each point.

The position reconstruction, shown in (B), is already fairly accurate thanks to the cross-shaped design of the metal electrodes. The largest migration is concentrated in the corners and near the metal arms: in these regions, the reconstruction clusters the points toward the closest read-out pad. 

\subsubsection{Use of signal area or Signal amplitude,  SWP or DPC}
 The amplitude of the signal is obtained by fitting a gaussian to the three or four highest samples around the signal peak, while the area is obtained by summing the areas of these bins. Since the clock in the digitizer is not synchronized to the laser trigger, these highest samples are not at fixed positions with respect to the signal peak. This fact introduces a large uncertainty in the determination of the signal area preventing   further its use in the analysis. For this reason, in the following part of this paper, only the signal amplitude  is used with both the DPC and SWP methods. 

\begin{figure}[htb]
\begin{center}
\includegraphics[width=0.95\textwidth]{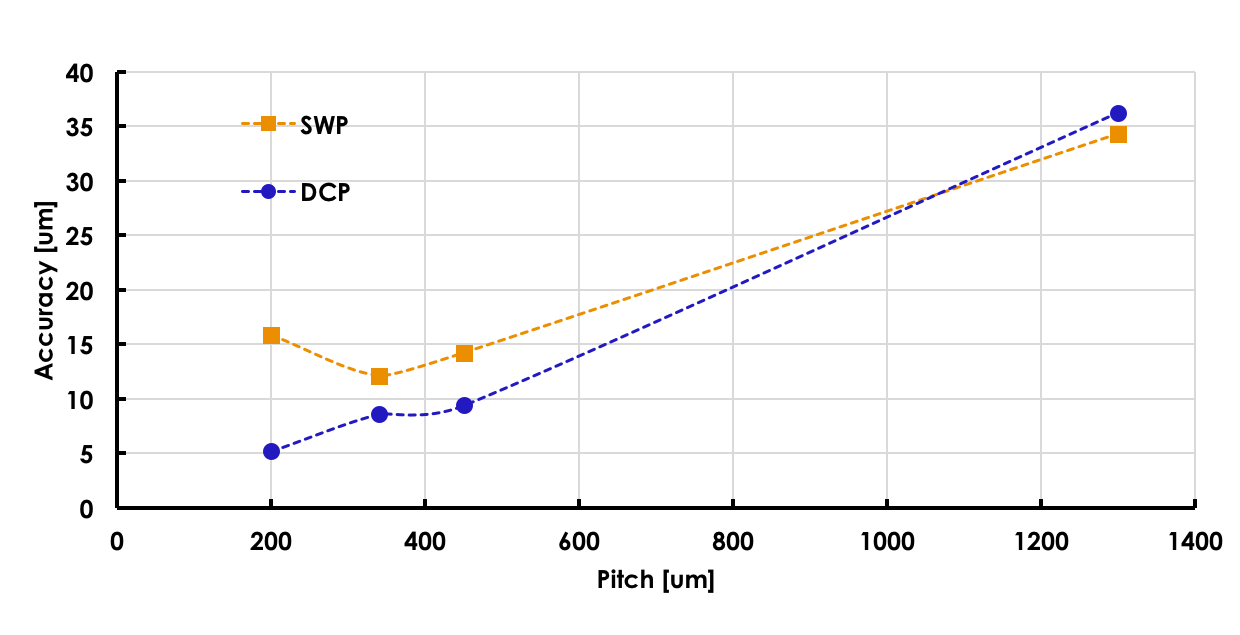}
\captionof{figure}{The measured accuracies for the two reconstruction methods  (DPC and SWP) as a function of the pitch size. }
\label{fig:recalgo}
\end{center}
\end{figure}

Figure~\ref{fig:recalgo} reports the measurement accuracies, defined as the mean difference between the true and reconstructed positions over the whole DUT, of the two reconstruction methods, as a function of the pixel size. Thanks to the possibility of tuning the $k_{x,y}$ parameters, the DCP method yields better results. For the largest pitch, 1300 \micron, the two methods have similar behavior since the best results are obtained for $k_{x,y} \sim 1$. As the pixel pitch gets smaller, the difference between the two methods grows, with SWP faring considerably worse for the smallest pitch. 

In the DPC algorithm,  the values $k_{x,y}$ = 0.6, 0.9, 0.85, 0.98 have been used for the pitch size 200, 340, 450, and 1300 \micron, respectively. 
For the above reasons, in the following of this analysis, the DCP algorithm with signal amplitude will be used. 

\subsubsection{Determination of the reconstructed coordinates}

As seen in Figure~\ref{fig:data}(C), the measured coordinates are systematically shifted with respect to their true positions. For a given hit position, this shift can be estimated by comparing the measured and true coordinates in the training dataset for those events whose reconstructed coordinates are in proximity (within a circle of $r_{cor}$)  to that of the event under study.  

\begin{equation} 
\begin{aligned}
\label{eq:kdpc2}
w_i = \frac{1.}{\sqrt{ (x_{meas} - x_{meas \;training}^i)^2 + (y_{meas} - y_{meas \; training}^i)^2}},\\
\Delta x = \frac{\sum_i^n (x_{meas \;training}^i - x_{true \;training}^i)\times w_i}{\sum_i^n w_i}, \\
\Delta y = \frac{\sum_i^n (y_{meas \;training}^i - y_{true \;training}^i)\times w_i}{\sum_i^n w_i}
\end{aligned}
\end{equation}
where $(x,y)_{true \;training}^i, (x,y)_{meas \;training}^i$ are respectively the true and measured $x,y$ coordinates of the $i$ training point.
The value of $r_{cor}$ does not have a strong impact on the correction, provided it is large enough to include at least a few training positions and not too large to include points that have different migration characteristics. For the present study, $r_{cor}$ was set to $r_{cor}$ = 30 \micron.  Once $\Delta x, \Delta y$ have been computed, the reconstructed hit coordinates are obtained as:

\begin{equation} 
\begin{aligned}
\label{eq:shift}
x_{rec} = x_{meas}  + \Delta x \\
y_{rec} = y_{meas}  + \Delta y \\
\end{aligned}
\end{equation}

\subsection{Reconstruction of the hit time}

The first significant difference in determining the hit time between RSD and standard UFSD is that in the RSD case, the time measured by a given electrode $i$, $t^i_{meas}$, is later than the hit time due to the delay, $t^i_{delay}$, introduced by the signal propagation on the resistive layer. 
Therefore, the reconstructed hit time $t^i_{rec}$ can be expressed as: 
\begin{equation}
\label{eq:timetrue}
t^i_{rec} = t^i_{meas} +t^i_{delay}+t^i_{setup},
\end{equation}

where $ t^i_{setup}$ is a hardware-specific offset due to PCB traces and cable lengths.

\begin{figure}[htb]
\begin{center}
\includegraphics[width=1\textwidth]{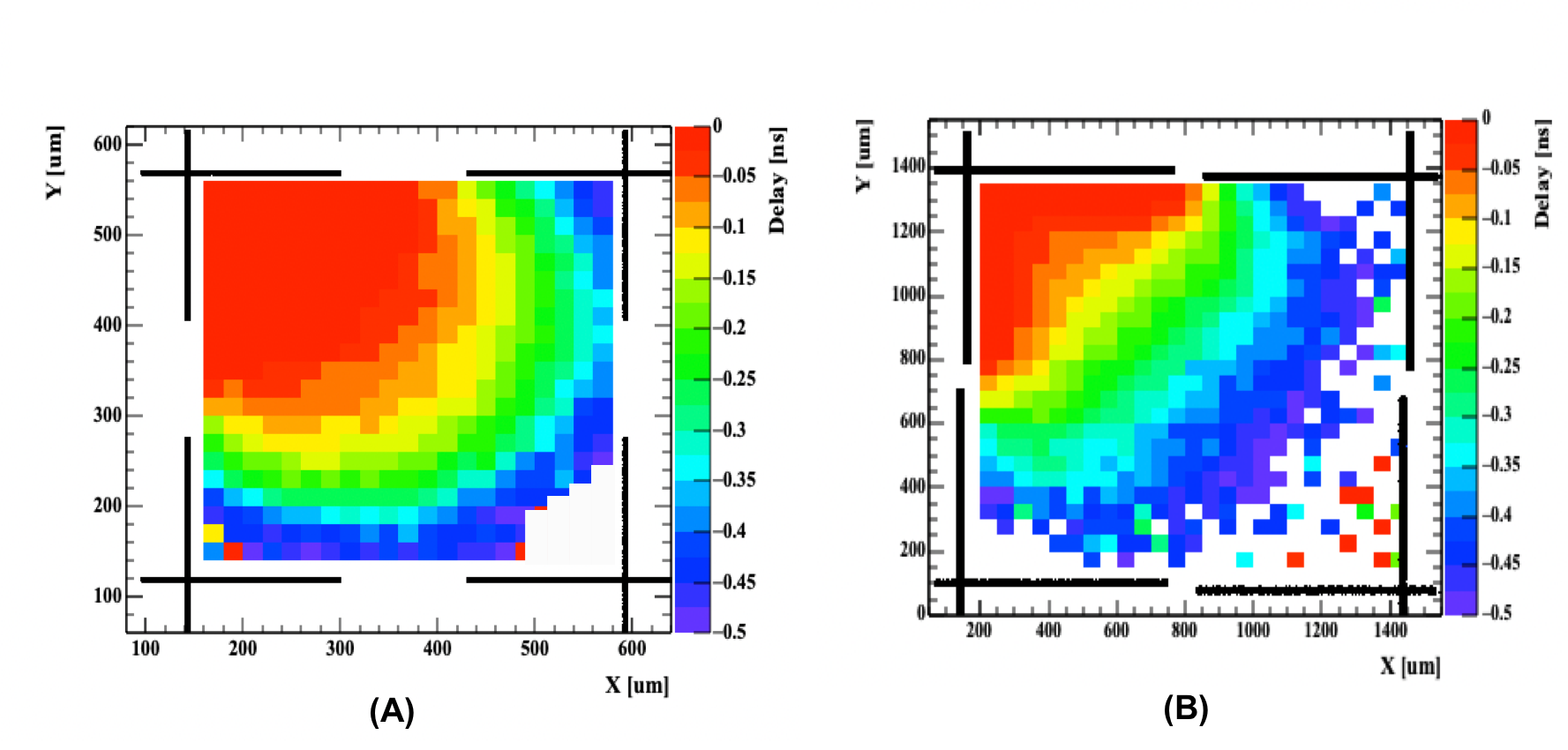}
\captionof{figure}{Signal delay  with respect to the top-left read-out pad (A) 450 $\times$ 450 \micronsqs structure.  (B)  1300 $\times$ 1300 \micronsqs structure. }
\label{fig:delay}
\end{center}
\end{figure}

In ~\cite{tornago2020resistive}, the delay has been measured to be about $t_{delay} \sim$ 0.3 - 0.5 ps/\micron, dependent on the surface resistivity and sensor capacitance.  Given the large pixel sizes used in this analysis, the delay can be as large as 300-400 ps. Figure~\ref{fig:delay}  shows the delay maps for the 450 $\times$ 450 \micronsqs and 1300 $\times$ 1300 \micronsqs structures as measured using the TCT set-up. In these plots, the signal is read out by the top-left electrode, and the colors illustrate the delay. For the 1300 \micronsqs structure, when the hit position is near the opposite corner, the signal amplitude is too small to allow determining the arrival time. 

The term  $t^i_{setup}$ is evaluated experimentally by measuring for each pad $i$ the time of arrival of laser signals shot very near the pad itself.

The second important difference between RSDs and standard UFSDs is that in RSDs there are multiple measurements of the hit time (one from each read-out pads), and their combination might improve (or deteriorate) the time resolution. The $\sigma_{Landau}$ does not benefit from multiple measurements, as the signal shape is common to all pads, while the jitter term does. 
The time of arrival $t_{rec}$ is estimated using the following $\chi^2$ expression:

\begin{equation}
\begin{aligned}
\label{eq:chi2}
\chi^2 = \frac{\sum_i^4(t_{rec} - t^i_{rec})^2} { \sum_i^4 \sigma_i^2},\\
\sigma_i = \frac{\sigma_{el-noise}}{dV_i/dt}  \sim      \frac{\sigma_{el-noise}} { A_i/t_{rise} },\\
\end{aligned}
\end{equation}
where $t^i_{rec}$, $A_i$, and $\sigma_i$  are the reconstructed hit time, the signal amplitude, and the time jitter measured on pad $i$, and $t_{rise}$ the signal rise time.
Minimizing the $\chi^2$ expression, and dividing out the common factors ($t_{rise}, \sigma_i$), the expressions for $t_{rec}$ and its associated error $\sigma_{trec}$ are found to be:

\begin{equation}
\begin{aligned}
\label{eq:thit}
 t_{rec} = \frac{\sum_i^4 t^i_{rec}*A^2_i} { \sum_i^4 A^2_i},\\
\sigma_{trec} =  \frac{\sigma_{el-noise} * t_{rise} } { \sqrt{\sum_i^4 A^2_i}}.\\
\end{aligned}
\end{equation}

Assuming for simplicity an equal signal split among the four pads, $A_i = A/4$, the expression for the error becomes $\sigma_{trec} =  \frac{\sigma_{el-noise} * t_{rise} \sqrt{4} } {A} $,  showing that also the time resolution worsens with the number of electrodes $n$ as $\sqrt{n}$.

\section{Sensors under study}
\label{sec:sus}
\begin{figure}[htb]
\begin{center}
\includegraphics[width=0.9\textwidth]{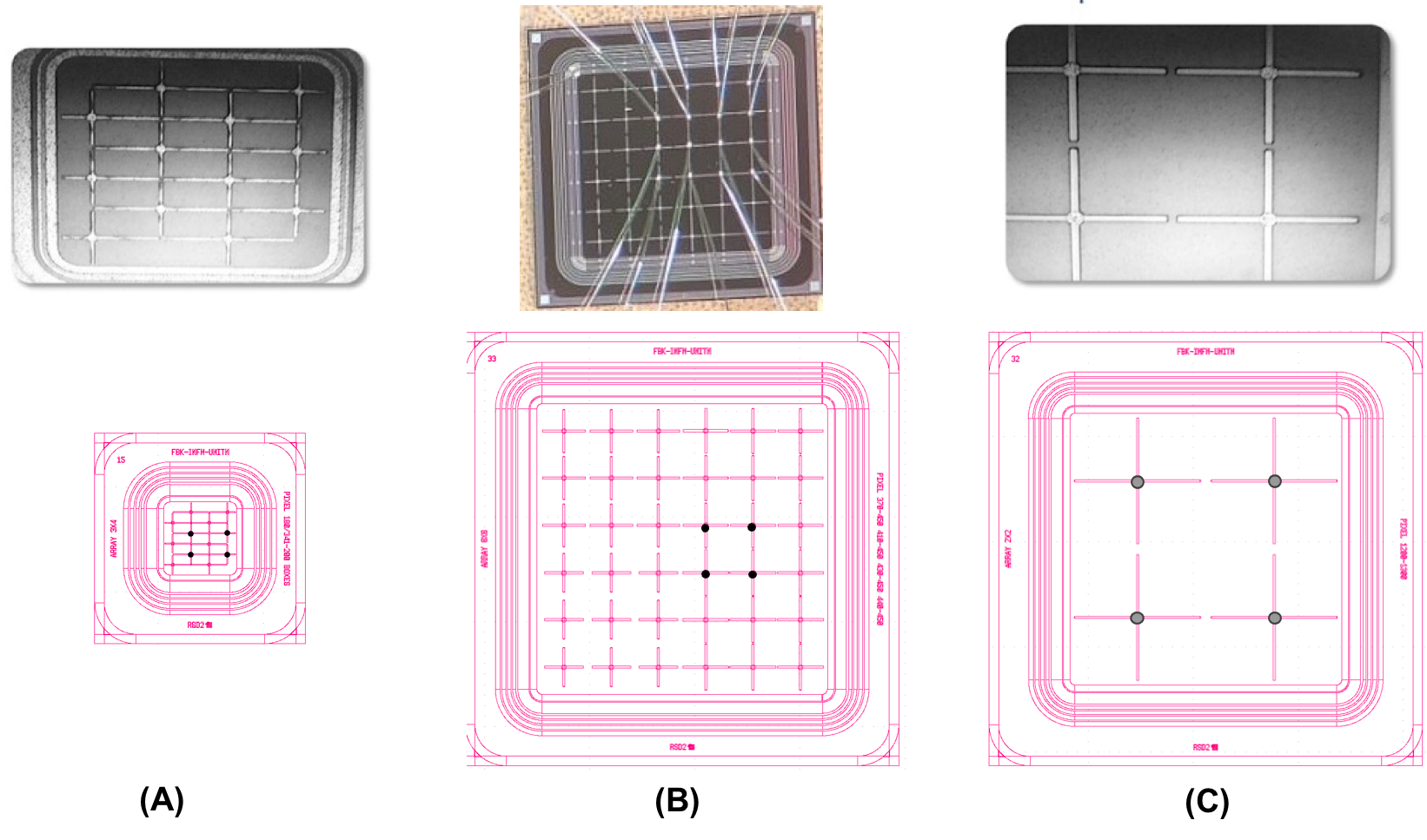}
\caption{Structures used in this study. In the bottom row,  the pictures are scaled, maintaining the original proportions. The read-out pads delimiting the pixel areas are indicated by the full dots. }
\label{fig:struct}
\end{center}
\end{figure}

The structures used in this study are shown in Figure~\ref{fig:struct} and their characteristics are listed in Table~\ref{tab:struc}. The only electrodes read out during the measurements are indicated with full dots (while the other electrodes are connected to ground).  The smallest sensor, (A), has an active area of 800 $\times$ 800 \micronsq, and has electrodes with arms of different lengths in the x and y directions, 90 \microns in x  and 165 \microns in y. A rectangular pixel of  200 $\times$ 345 \micronsqs is obtained by leaving the electrode internal to the four read-out pads floating. The structures (B) and (C) have an active area of about  2700 $\times$ 2700 \micronsq. The type (B) has a 6 x 6 array of read-out electrodes, with a pitch of 450 \micron, while (C) as four read-out pads, defining a single pixel with a pitch of 1300 \micron. 

\begin{figure}[tbhp]
\begin{center}
\includegraphics[width=1.\textwidth]{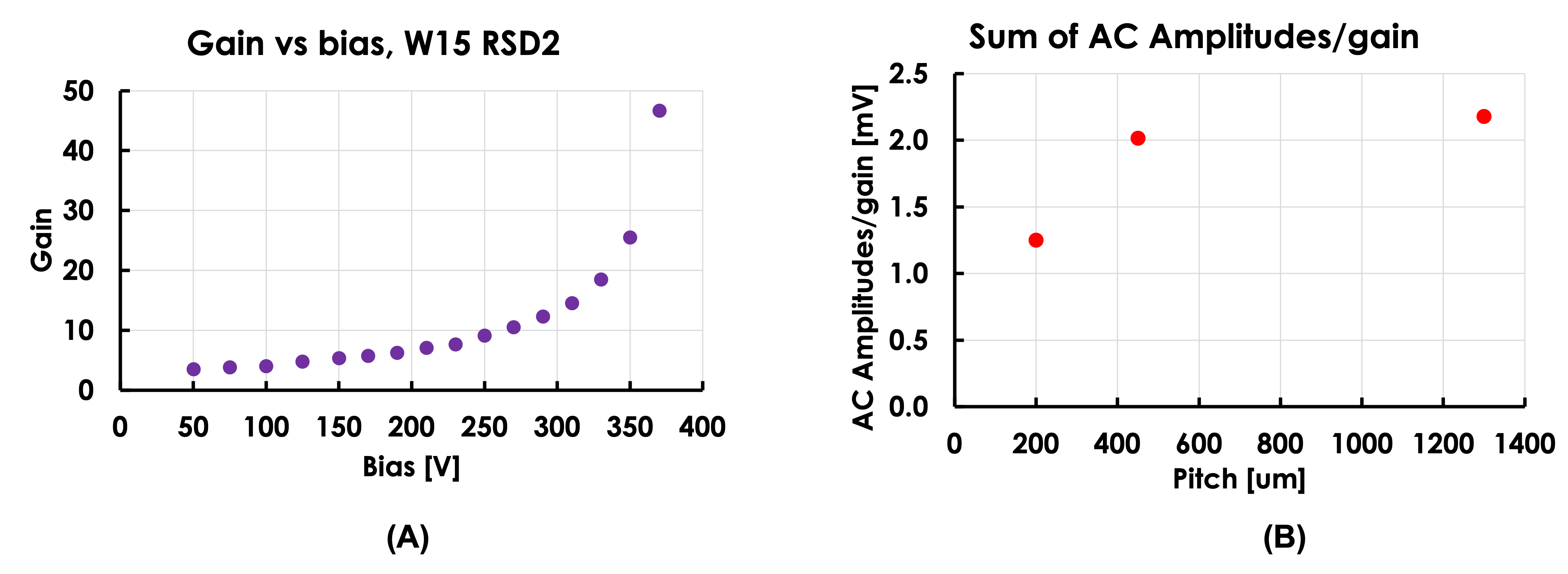}
\captionof{figure}{(A): Gain-voltage characteristics of the sensors used in this analysis. (B): Sum of the four amplitudes of the AC signals divided by the gain. In small pixels, the AC signal is not contained within the four closest read-out pads, so the fraction is lower. }
\label{fig:biasV}
\end{center}
\end{figure}

The structures have been selected from the same wafer, so they have the same n$^+$ sheet resistivity and gain vs. bias behaviour, shown in Figure~\ref{fig:biasV}(A). In order to study how the various components of the spatial resolution evolve over a wide signal range, lasers with intensities higher than 1 MIP have been used. For this reason, in the pictures, the gain is reported as "equivalent gain", meaning the product of the gain and the laser setting, expressed in 1-MIP unit. 

 Figure~\ref{fig:biasV}(B) reports the sum of the amplitudes measured on the four read-out pads divided by the area of the signal measured on the $n^+$ resistive layer (in the following called DC-signal) as a function of the pitch. In large structures, 450 \microns and 1300 \microns pitch, the AC signal is fully contained by the four read-out pads, and the ratio does not depend on the pitch.  On the other hand, for the 200 $\times$ 340 \micronsqs structure, this ratio is about 40\% lower since the signal sharing also involves neighboring pixels, and the signal is not limited to the four  closest read-out electrodes.  As this analysis uses only four read-out electrodes, the resolution for this smaller structure is degraded. 
 
 This observation  highlights an important interplay between the $n^+$ resistivity, the pixel size, and the optimal number of read-out electrodes needed to reconstruct the signal: 
in order to contain signal sharing to the four electrodes at the pixel corners, the $n^+$ resistivity should be tuned according to the pixel size, it should be lower for larger pixels and higher for smaller pixels. 

\subsection{Alignment and signal shape}
\begin{figure}[tbhp]
\begin{center}
\includegraphics[width=0.9\textwidth]{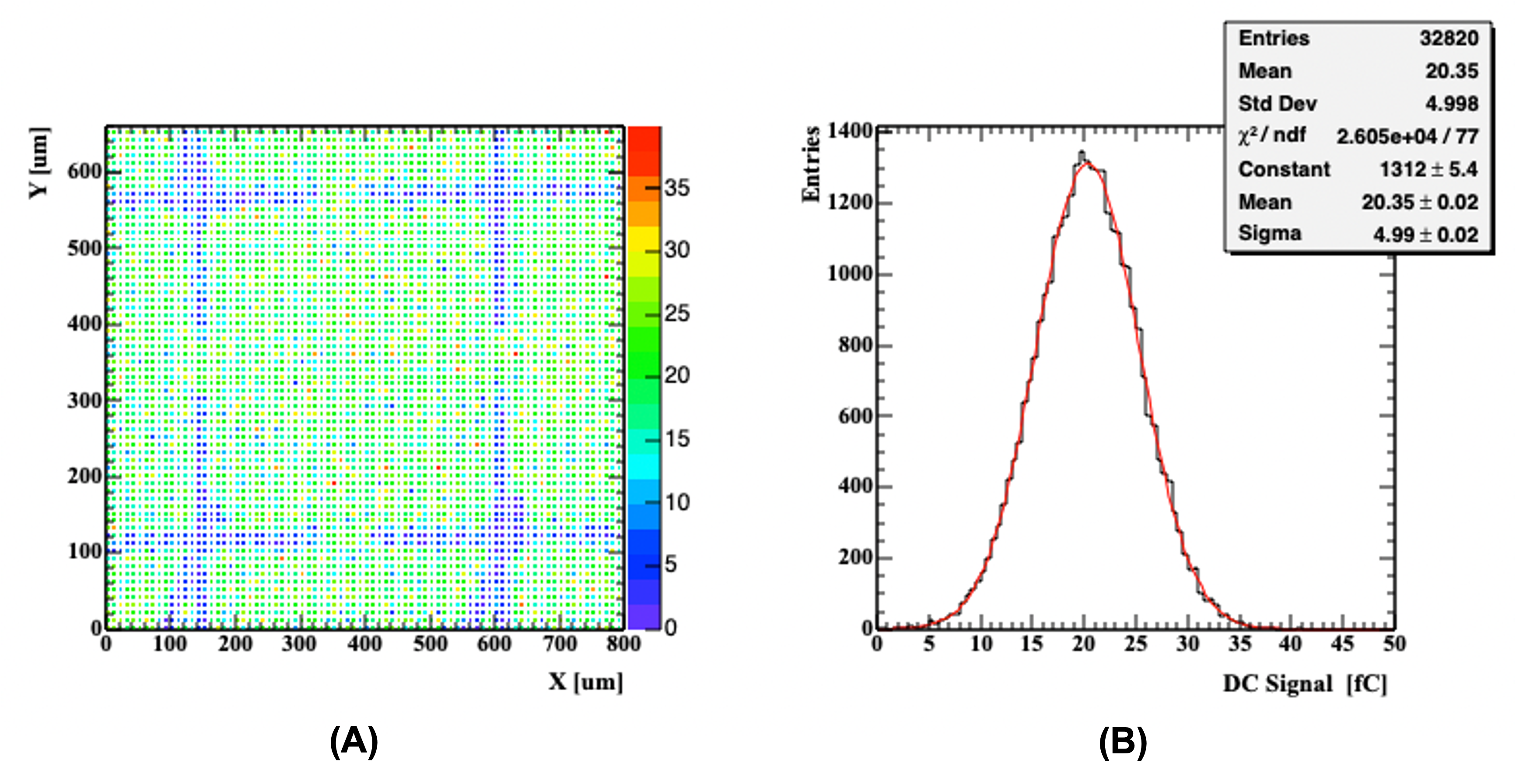}
\captionof{figure}{(A): Area of the DC-signal in fC as a function of position. (B): 1D distribution of the signal charge for shot inside the pixel. }
\label{fig:DC450}
\end{center}
\end{figure}
\begin{figure}[h!]
\begin{center}
\includegraphics[width=0.8\textwidth]{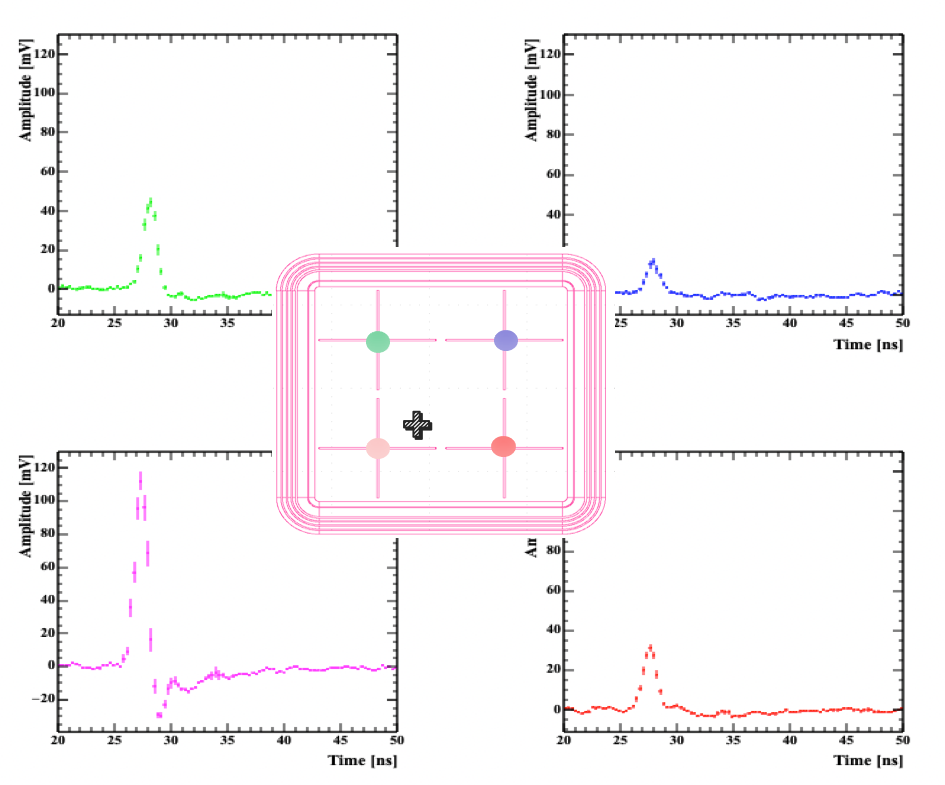}
\captionof{figure}{AC signals  on the four read-out electrodes for the 1300 \microns pixel structure when the laser is shot at the position indicated by the cross }
\label{fig:shape}
\end{center}
\end{figure}

For each structure under test, the first step is to find the pads coordinates in the laser reference system. This is done by exploiting the fact that the metal of the read-out pads absorbs the laser signal: Figure~\ref{fig:DC450}(A) shows the DC-signal area, in fC, as a function of the laser position for a pixel of 450 \micron. The image clearly shows the metal arms of each read-out pad. For the lower two read-out electrodes, the wire bonds are also visible. Figure~\ref{fig:DC450}(B) reports the  1D distribution of the signal charge for the shots inside the pixel. The plot has a very regular gaussian shape, without long tails.
Figure~\ref{fig:shape} shows the AC signals on the four read-out electrodes for the 1300 \microns pixel structure when the laser is shot at the position indicated by the cross.  

The signals are very fast, about 2 ns long, and are not distorted even by a rather long propagation, about 1 mm for the green, blue, and red signals. The opposite polarity lobe of the signal is quite small, indicating a fairly long RC time constant. More details on signal propagation and the evaluation of the RC time constant can be found in ~\cite{tornago2020resistive},

\section{Evaluation of the spatial resolution}
\label{sec:space}
In the following, the spatial resolution for a given device is estimated as a function of the gain.  As the migration matrix has been measured on the device under test, the terms $\sigma_{setup}$ and $\sigma_{sensor}$ of Equation~\ref{eq:spaceres} are by construction equal to zero.  An estimate of these two terms is provided in section~\ref{sec:terms}.

\begin{figure}[thbp]
\begin{center}
\includegraphics[width=1.\textwidth]{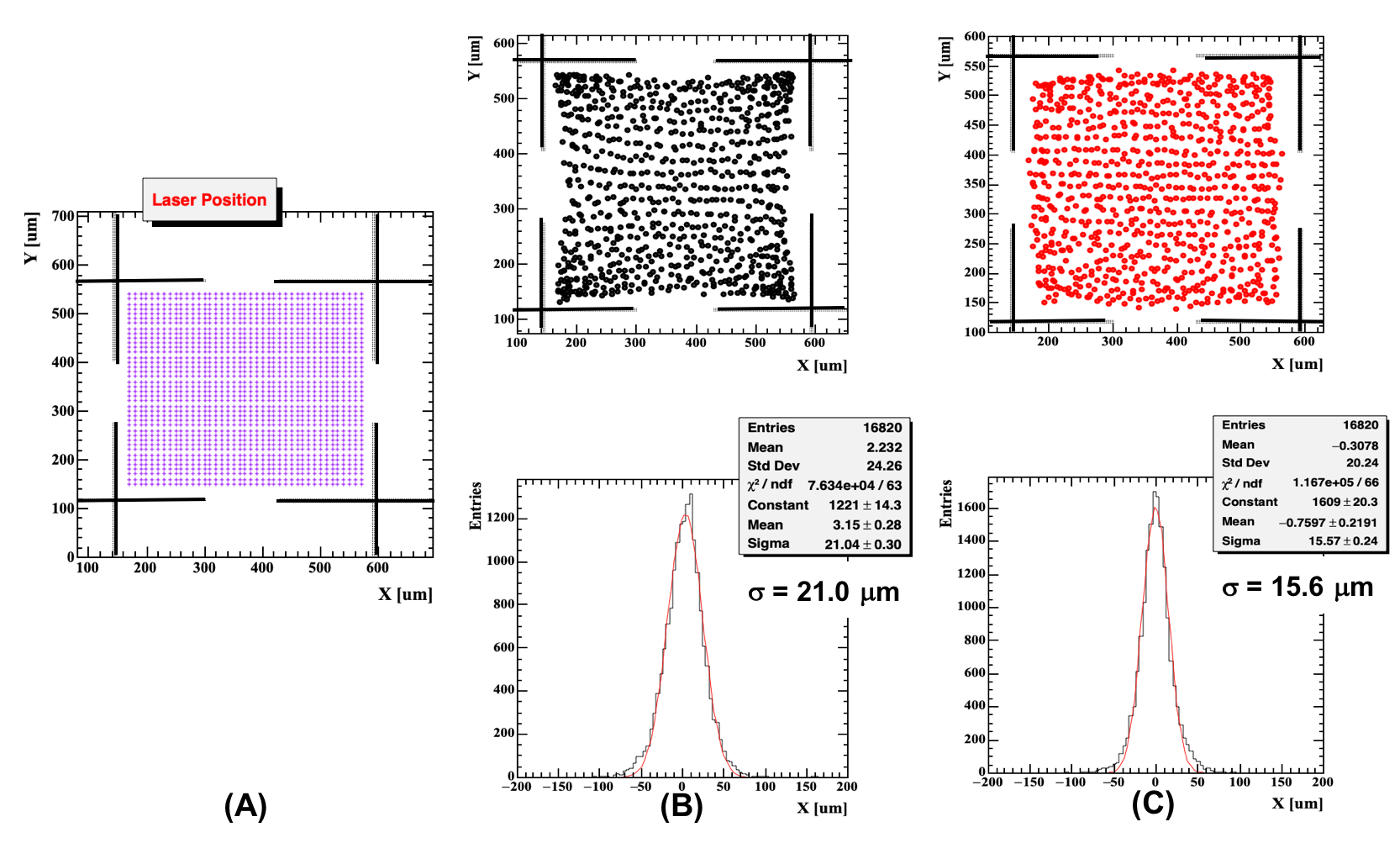}
\captionof{figure}{(A) Position of the laser shots, (B) Uncorrected position reconstruction, (C) Corrected position reconstruction}
\label{fig:cor}
\end{center}
\end{figure}

For each sensor, at every biasing point,  the following steps are performed:
\begin{itemize}
\item The laser is shot in a grid of points covering the pixel area. The step size is 10 \microns for the 200 $\times$ 340 \micronsqs and 450 $\times$ 450  \micronsqs structures, while it is 20 \microns for the 1300 $\times$ 1300  \micronsqs structure. This is illustrated on Figure~\ref{fig:cor}(A). 
\item The hit positions are reconstructed using Equation~\ref{eq:dpc}: Figure~\ref{fig:cor}(B upper plot) shows  these reconstructed positions for the  450 $\times$ 450  \micronsqs structure. Thanks to the read-out electrode design, the resolution,  reported in Figure~\ref{fig:cor}(B bottom plot),  is quite good, $\sigma_{hit\; pos} $  = 21.0 \microns.  Small non-gaussian tails are visible due to the clustering of the reconstructed positions near the electrodes.  
\item  The reconstructed positions are corrected using the procedure outlined in Section~\ref{sec:rec}. The position of the laser shots is required to be at least 30 \microns away from the metal strips of the read-out pads in order to assure that the laser has not been inadvertently attenuated.   
The effect of the correction can be gauged by comparing Figure~\ref{fig:cor}(B) and (C): the distortion in the reconstruction is almost completely eliminated, and the corrected points form a more regular grid. The position resolution improves, from  $\sigma_{hit\; pos} $ = 21.0 \microns to $\sigma_{hit\; pos} $  = 15.6 \micron, since the accuracy of the reconstruction becomes much better (smaller $\sigma_{rec} )$  and the terms $\sigma_{setup}$ and $\sigma_{sensor}$ are eliminated by the correction. 
\end{itemize}

\begin{figure}[th!]
\begin{center}
\includegraphics[width=0.95\textwidth]{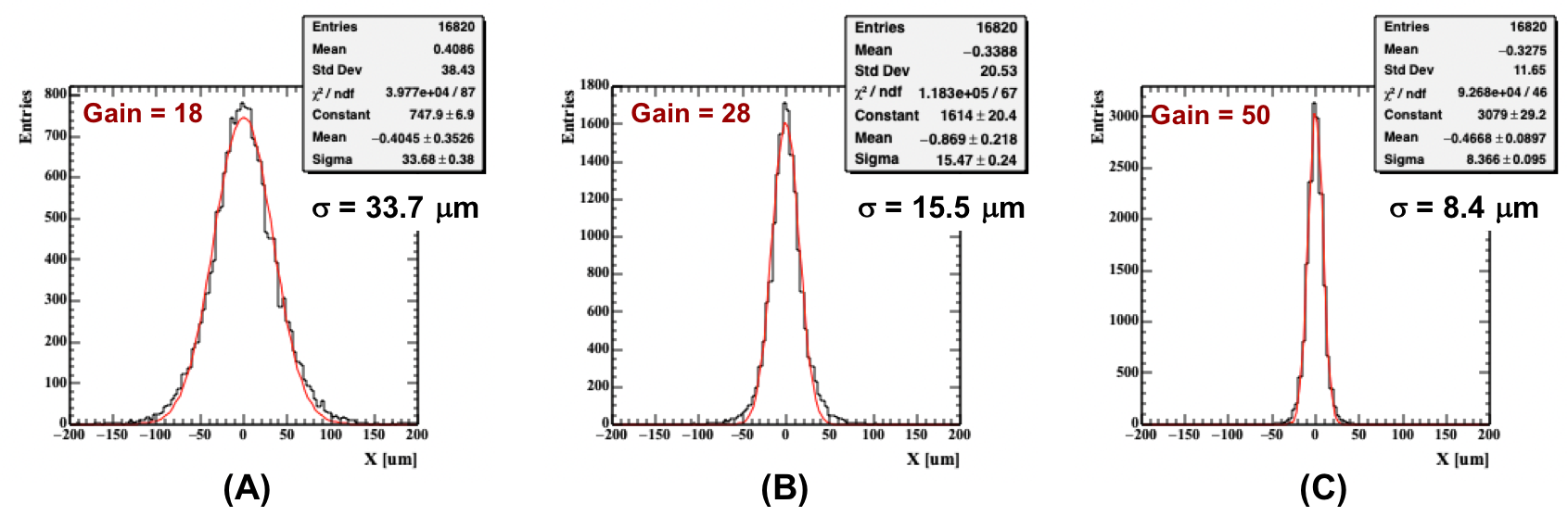}
\captionof{figure}{Spatial resolution as a function of the RSD gain for the  450 $\times$ 450  \micronsqs structure. In these measurements,  the two terms $\sigma_{setup},\sigma_{sensor}$ are zero.  }
\label{fig:resgain}
\end{center}
\end{figure}

The evolution of the spatial resolution with gain for the   450 $\times$ 450  \micronsqs structure is shown in Figure~\ref{fig:resgain}. Gain 18 (A) and 28 (B) were obtained with the laser set at one MIP, while, for gain 50 (C), the laser was set to about two MIPs. For all values of gain, the non-gaussian tails are very small, indicating that the correction procedure works correctly.

\subsection{Results}
\begin{figure}[tbp]
\begin{center}
\includegraphics[width=1.\textwidth]{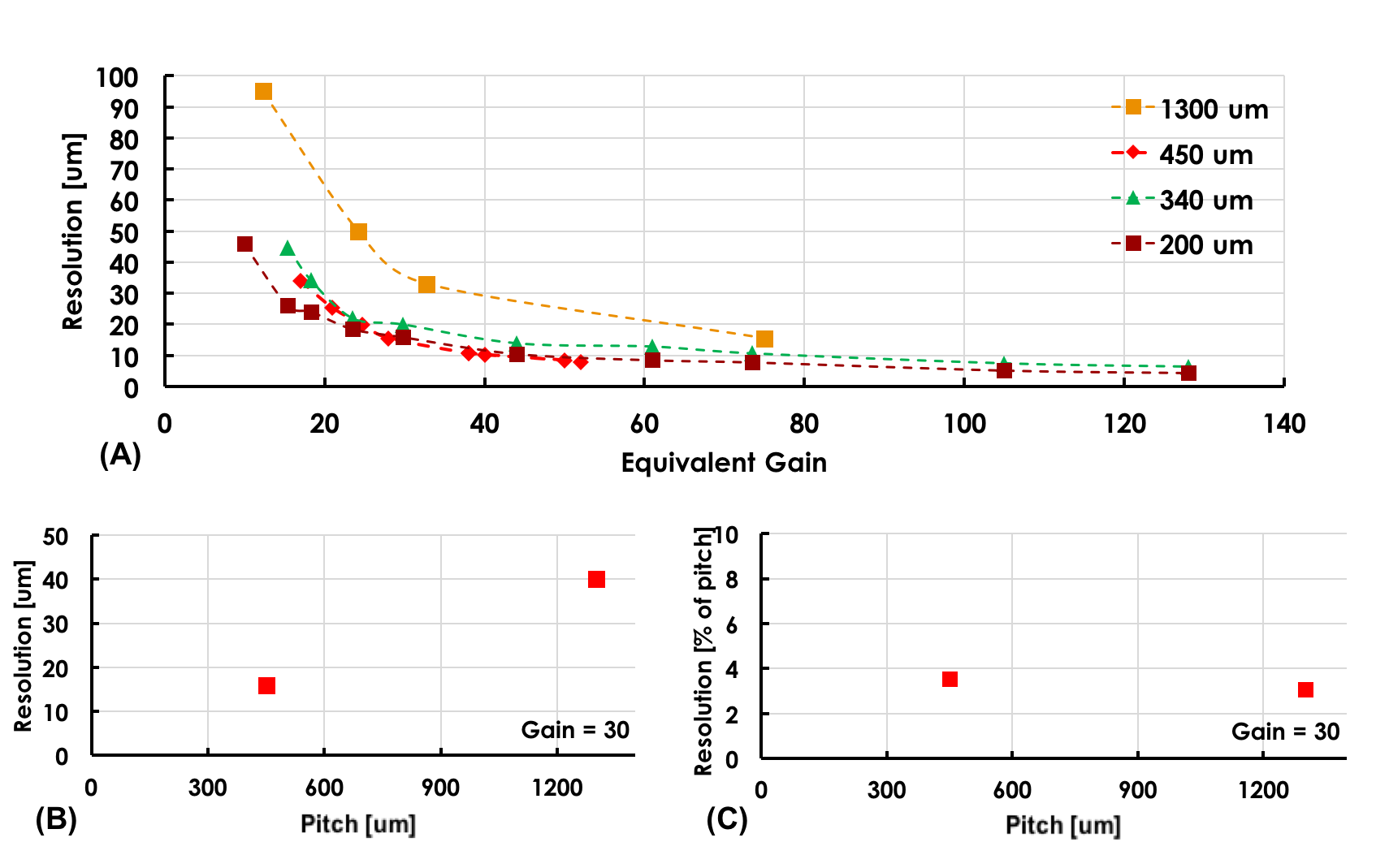}
\captionof{figure}{ (A) The spatial resolution versus gain for the four pitch sizes analyzed.  For the two largest structures: (B) Spatial resolution vs. the pitch size, (C)  Spatial resolution, expressed as a percentage of the pitch size vs. pitch size. In these measurements,  the two terms $\sigma_{setup},\sigma_{sensor}$ are zero. 
}
\label{fig:res}
\end{center}
\end{figure}

The spatial resolution as a function of gain for the different sensor types is presented in Figure~\ref{fig:res}(A). The resolutions for the 200 and 340 \microns pitches are not as good as they could be since, as anticipated in section~\ref{sec:sus}, the signal is not contained in the four  read-out pads.  For the two largest structures, at gain 30, Figure~\ref{fig:res}(B) reports the spatial resolution versus the pitch size, while Figure~\ref{fig:res}(C)  expresses the spatial resolution as a percentage of the pitch. A spatial resolution of about 3\% of the pitch size is achieved at gain 30. 

\begin{figure}[htbp]
\begin{center}
\includegraphics[width=0.95\textwidth]{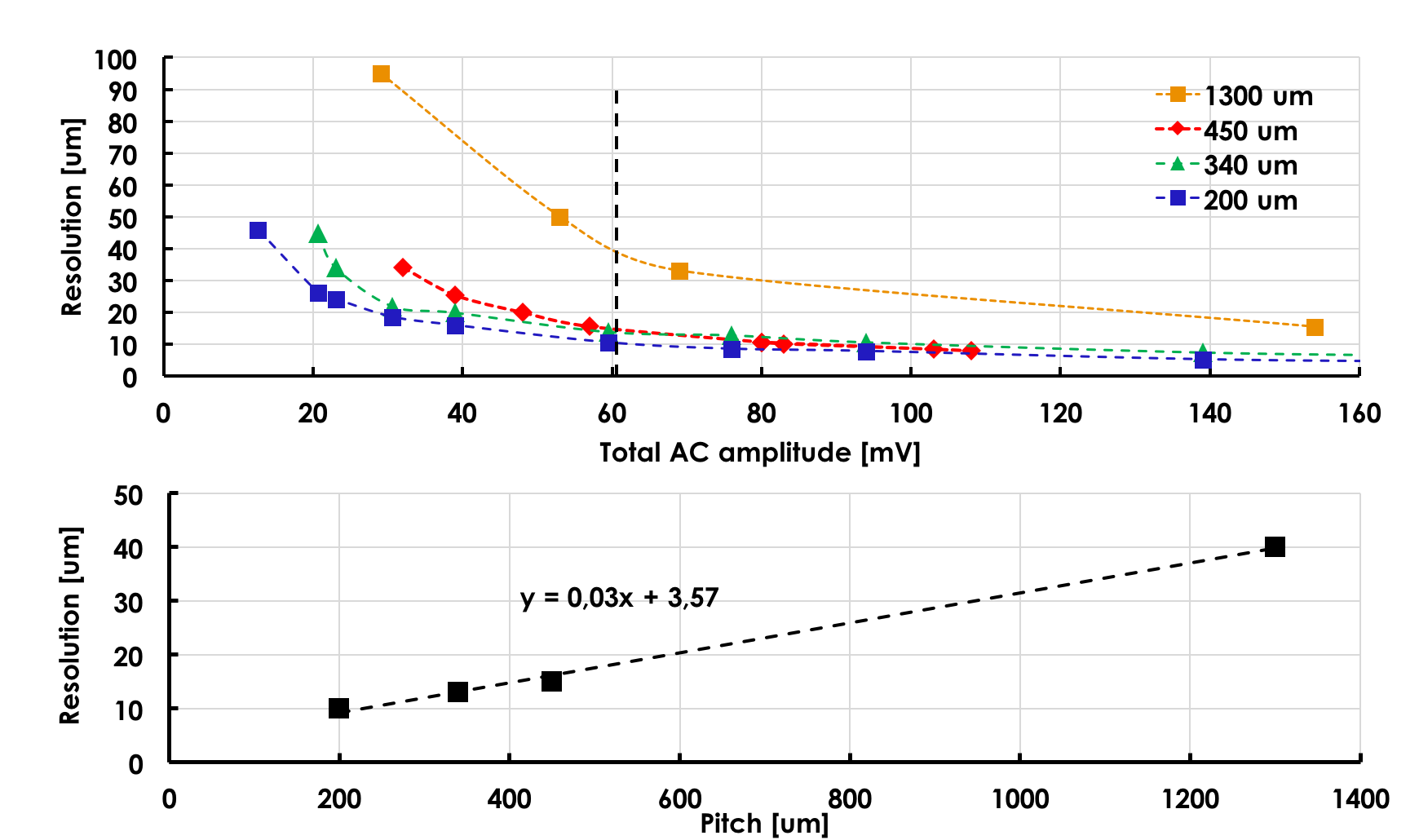}
\captionof{figure}{Top: Spatial resolution versus total amplitudes. Bottom:  Spatial resolution versus pitch size when the total AC amplitude equals 60 mV. As predicted by Equation~\ref{eq:jitterxx}, at a fixed amplitude, the resolution depends linearly on the pixel size.  In these measurements,  the two terms $\sigma_{setup},\sigma_{sensor}$ are zero.  }
\label{fig:resampl}
\end{center}
\end{figure}

The top plot in Figure~\ref{fig:resampl} shows the spatial resolution as a function of the total  AC amplitude, defined as the sum of the amplitudes measured on the four read-out electrodes. 
As expected, for equal signal amplitude, the smaller pitch sizes perform better. The bottom plot reports the spatial resolution as a function of the pitch at amplitude = 60 mV (about gain = 30). At fixed amplitude,  the resolution scales linearly with the pixel size, as it should happen when the resolution is dominated by the jitter, see Equation~\ref{eq:jitterxx},
The fit indicates a resolution of 3\% the pixel size with an offset of 3.5 \micron.

\begin{figure}[htbp]
\begin{center}
\includegraphics[width=0.95\textwidth]{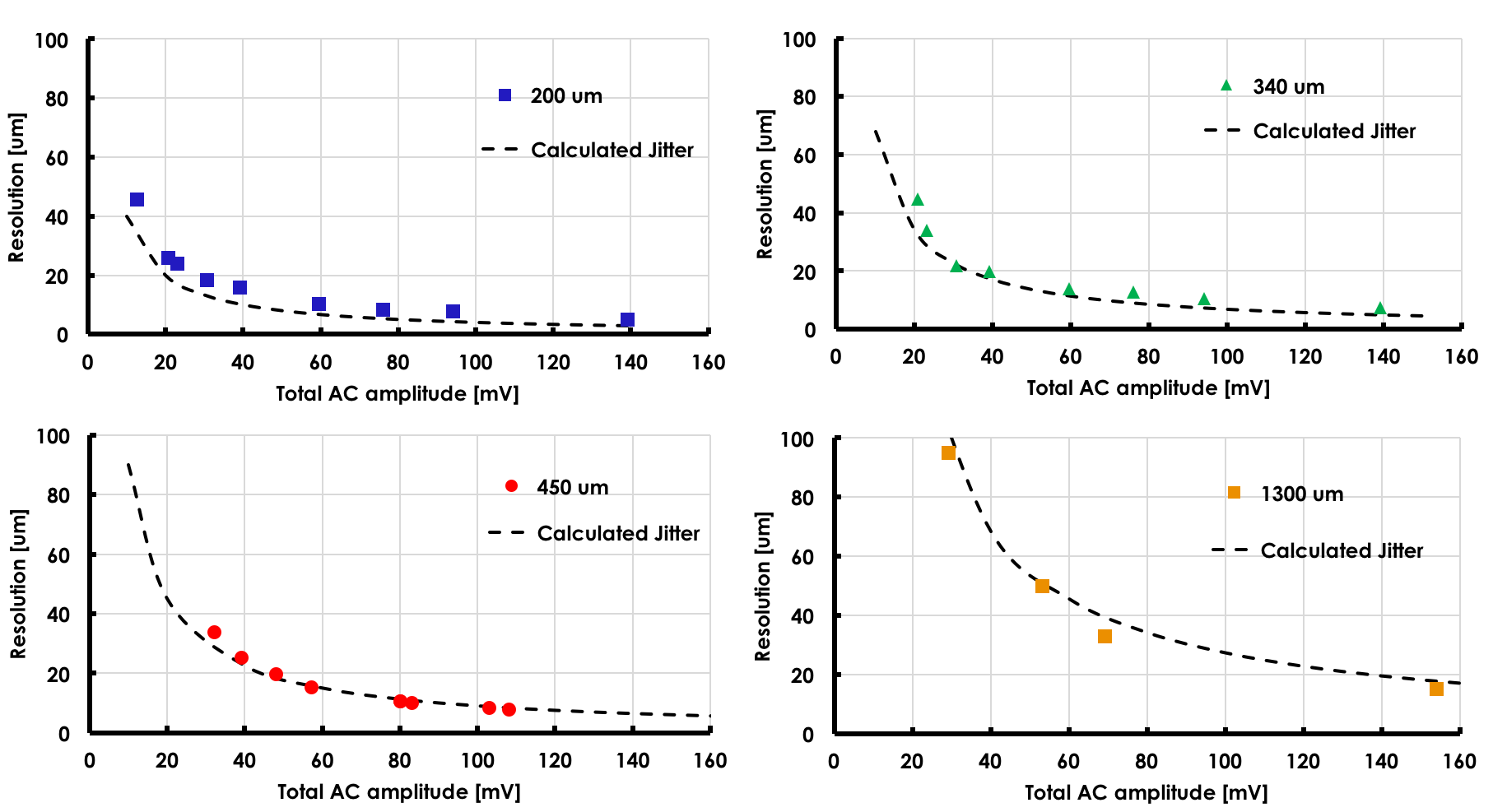}
\captionof{figure}{Spatial resolution vs. total AC amplitude for each pitch size and the calculated jitter contribution.  In these measurements,  the two terms $\sigma_{setup},\sigma_{sensor}$ are zero. }
\label{fig:fixedgain}
\end{center}
\end{figure}

Figure~\ref{fig:fixedgain} compares the resolution of each pitch size with the jitter contribution. Even though the only degree of freedom of the calculated jitter curves is a  common normalization parameter, the agreement with the experimental data is quite good.

\subsection{Evaluation of the  $\sigma_{setup}^2 +\sigma_{sensor}^2$ terms}
\label{sec:terms}
\begin{figure}[h]
\begin{center}
\includegraphics[width=0.95\textwidth]{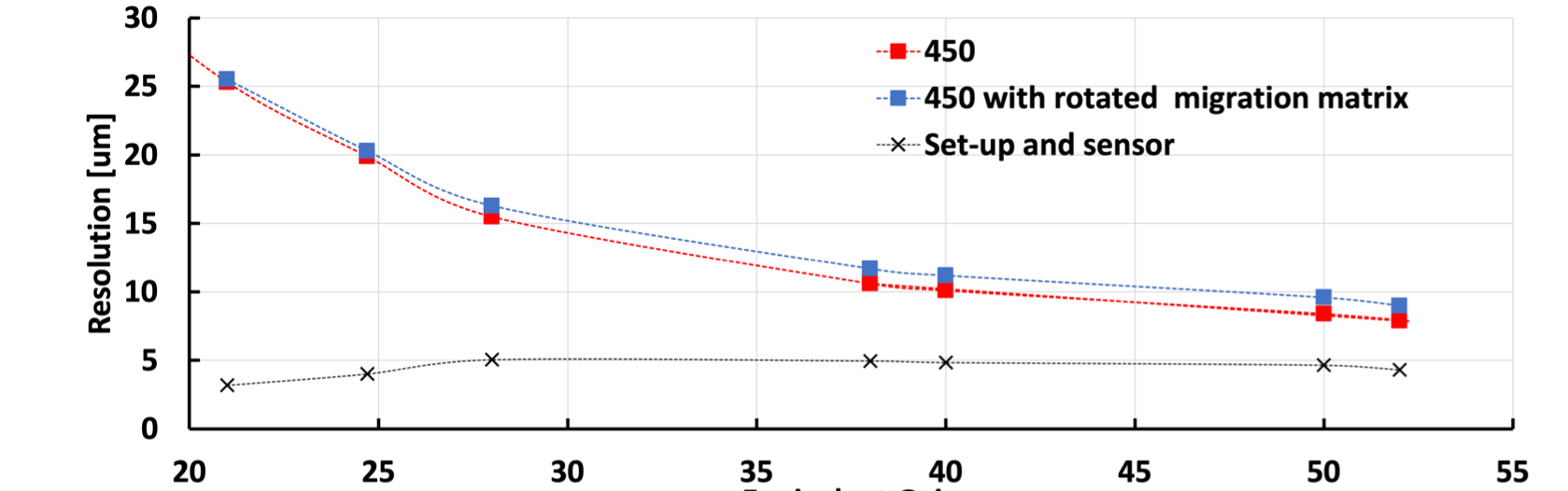}
\caption{Spatial resolution as a function of gain for the  450 $\times$ 450  \micronsqs structure. The red squares were obtained with the standard procedure, while for the empty blue squares the migration map was rotated by 180$^o$. The symbols $\times$ show the difference (in quadrature) between the blue and red squares}
\label{fig:system}
\end{center}
\end{figure}

In the results presented above, the migration matrix minimizes the term  $\sigma_{rec}$ and removes the combined contributions of  $\sigma_{setup}^2 +\sigma_{sensor}^2$ since it is computed on the same pixel used for the analysis. An estimate of  $\sigma_{setup}^2 +\sigma_{sensor}^2$  for the present study can be evaluated by rotating the migration map, for example, by 180$^o$. 
With this operation, possible differences arising from the TCT set-up, read-out amplifiers, and non-uniformity of the sensor sharing quality (for example, non-uniform resistivity or oxide thicknesses) are enhanced since the migration patterns of the points on the left (top) of the pixel center are applied to the points on the right (bottom) and vice versa. Figure~\ref{fig:system} shows the resolution vs. gain for the 450 \microns pitch pixel obtained by applying the standard and the 180$^o$ rotated migration matrix. Using the rotated migration matrix, the resolution is always slightly higher, and the difference in quadrature of the two resolutions is fairly constant and equal to about 5 \microns (shown in the picture with the symbols $\times$ connected with a black dotted line). 

The same analysis performed on the 1300 \microns structure leads to a value of $\sigma_{setup}^2 +\sigma_{sensor}^2 \sim$ 4.5 \micron, while on the smaller structure  the  use of the rotated migration matrix leads to values of  spatial resolution compatible with the standard matrix.

\section{Evaluation of the time resolution}
\label{sec:time}
This study was performed using the TCT set-up, measuring the difference between the trigger time and reconstructed event time $t_{rec}$ of laser shots distributed over the whole pixel surface.

\begin{figure}[tp]
\begin{center}
\includegraphics[width=1.0\textwidth]{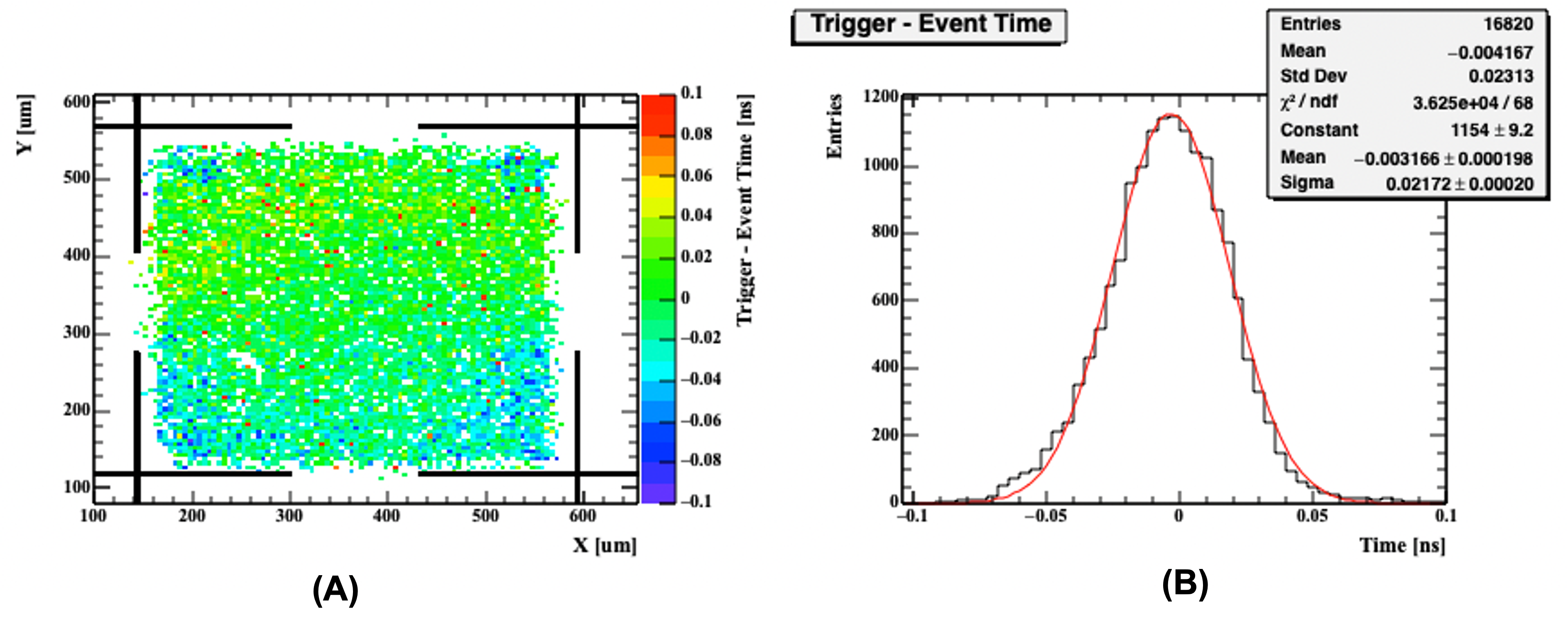}
\captionof{figure}{Time difference between the trigger and the  reconstructed event time (A) on the pixel surface  (B)  1D distribution.}
\label{fig:onept}
\end{center}
\end{figure}

Figure~\ref{fig:onept}(A) shows the  quantity   $t_{trig} - t_{rec}$ as a function of the hit position for the 450 $\times$ 450  \micronsqs structure, while (B) shows  the 1D distribution. 

The 2D map shows a very good uniformity over the whole surface, especially considering that this pixel is quite large.  This consideration is strengthened by the very small non-gaussian tail of the 1D distribution shown on (B). The events with the poorer resolution are concentrated near the pixel periphery.  
 In this analysis, the best results have been obtained using as $t^i_{meas}$ the time of the maximum and not the value obtained with the more common constant fraction algorithm. This feature is linked to the limited number of samples on the signal rising edge (the digitizer has a 5 GS/s sampling rate)  and not to a specific aspect of the resistive read-out.

Overall, these results show that resistive read-out does not degrade the timing performance of the UFSD design and that very uniform response over large pixels is achievable.

\subsection{Results}
\begin{figure}[htb]
\begin{center}
\includegraphics[width=0.9\textwidth]{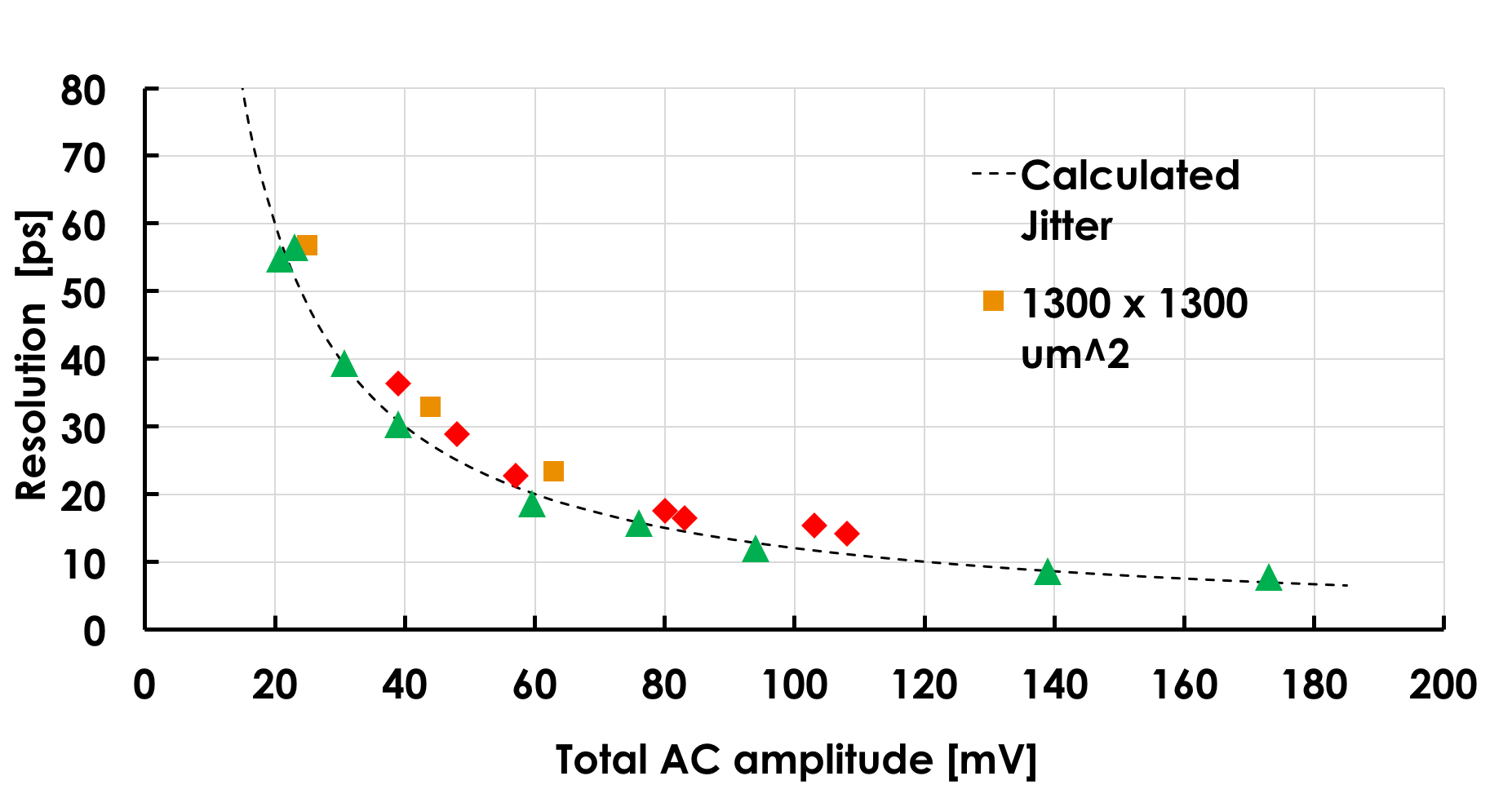}
\captionof{figure}{Time resolution as a function of the total AC amplitude. Results obtained with a laser TCT system ($\sigma_{Landau}$ = 0 ps). }
\label{fig:jitter}
\end{center}
\end{figure}

The complete set of measurements for the three RSD structures is shown in Figure~\ref{fig:jitter}.  The trigger resolution, evaluated at 10 ps, has been subtracted in quadrature. 
The resolution is presented as a function of the total AC amplitude.  It is important to stress the following points:
\begin{itemize}
\item Since a laser shot creates uniform charge deposition, in this study the term $\sigma_{Landau}$ is absent. This contribution has been measured~\cite{tornago2020resistive}  to be around 30 ps for a 50 \microns thick  RSD sensor. 
\item Given the excellent spatial resolution, the term $\sigma_{delay}$ is sub-leading with respect to $\sigma_{jitter}$.
\end{itemize} 
Therefore, the time resolution is dominated by the jitter contribution. One feature is particularly striking: the points align quite well along the curve representing the jitter contribution, regardless of the pixel size. This indicates that the jitter depends mostly upon the total AC amplitude, and it is not spoiled by the propagation on the n$^+$ resistive surface.  

Assuming to work at a total AC amplitude of 60 mV (gain = 30), a time resolution of about 19 ps is achieved for the 200 $\times$ 340 \micronsqs structure and 23 ps for the  450 $\times$ 450  \micronsqs and 1300 $\times$ 1300  \micronsqs structures.

\section{Extrapolated performance of RSD sensors with MIP.}
\begin{figure}[htb]
\begin{center}
\includegraphics[width=0.9\textwidth]{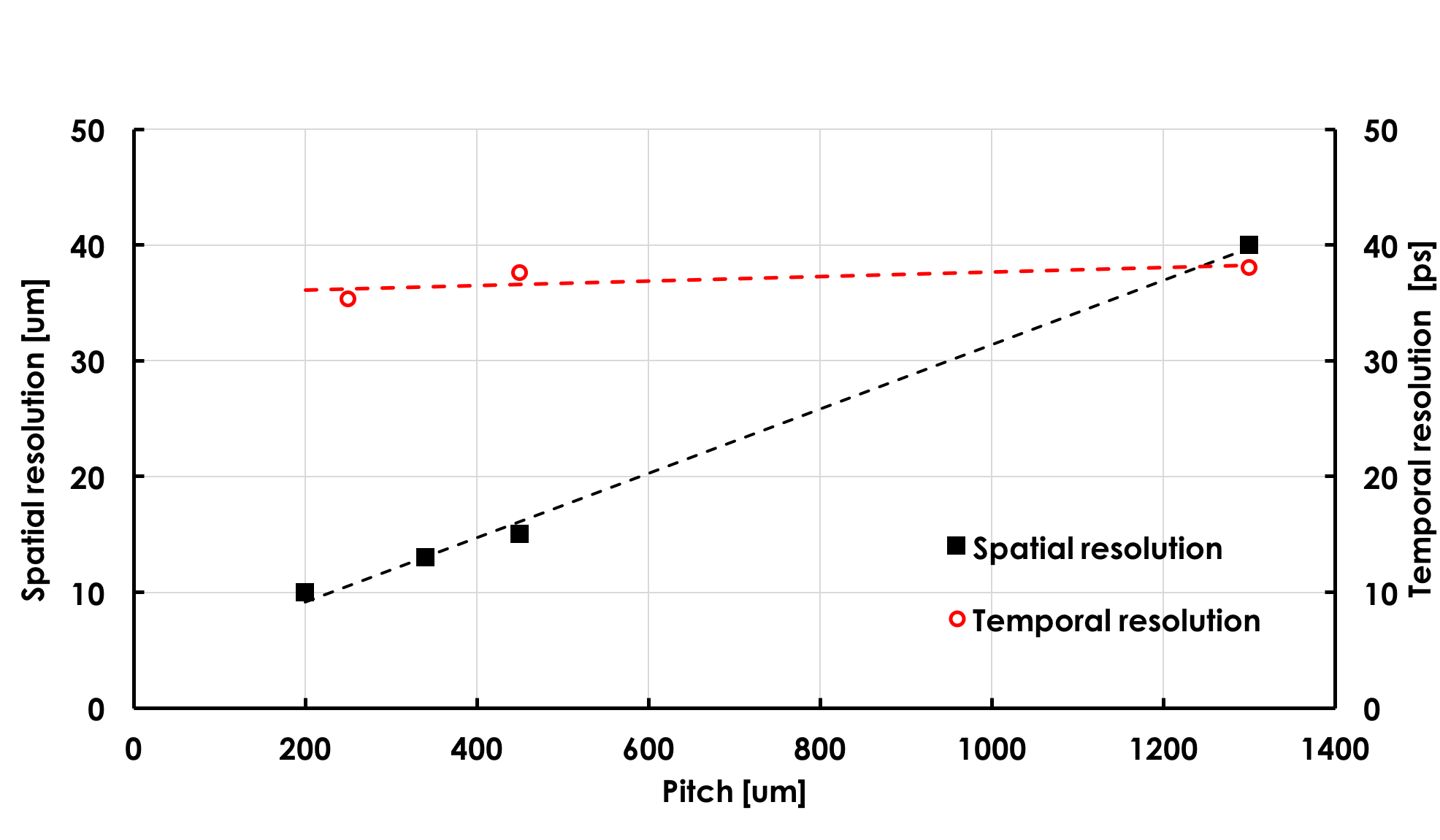}
\captionof{figure}{Space and time resolutions for the structures under test when the sum of the AC amplitudes is 60 mV (gain = 30).}
\label{fig:resolutions}
\end{center}
\end{figure}

The extrapolated resolutions for the determination of the position and time coordinates, for the sensors under test, are presented in Figure~\ref{fig:resolutions}.  The time resolution has been computed by adding  the Landau noise term ($\sigma_{Landa \; noise}$ = 30 ps) in quadrature to the time jitter term while the spatial resolution by adding $\sigma_{setup}^2 +\sigma_{sensor}^2$ = 5 \microns in quadrature to the spatial term of the 450 and 1300 \microns structures.

\begin{itemize}
\item The spatial resolution is about 3\% of the pixel size, and it scales linearly with the pixel size, as predicted by Equation~\ref{eq:spaceres}
\item The temporal resolution is fairly constant at about 38 ps as a function of the pixel size.
\end{itemize}

These results demonstrate that RSD sensors with cross-shaped electrodes are able to achieve excellent resolutions in the determination of the position and time coordinates, over a very large range of pixel sizes. 

\section{Resolution, occupancy, and power consumption for different RSD pixel shapes.}
\begin{figure}[htb]
\begin{center}
\includegraphics[width=0.9\textwidth]{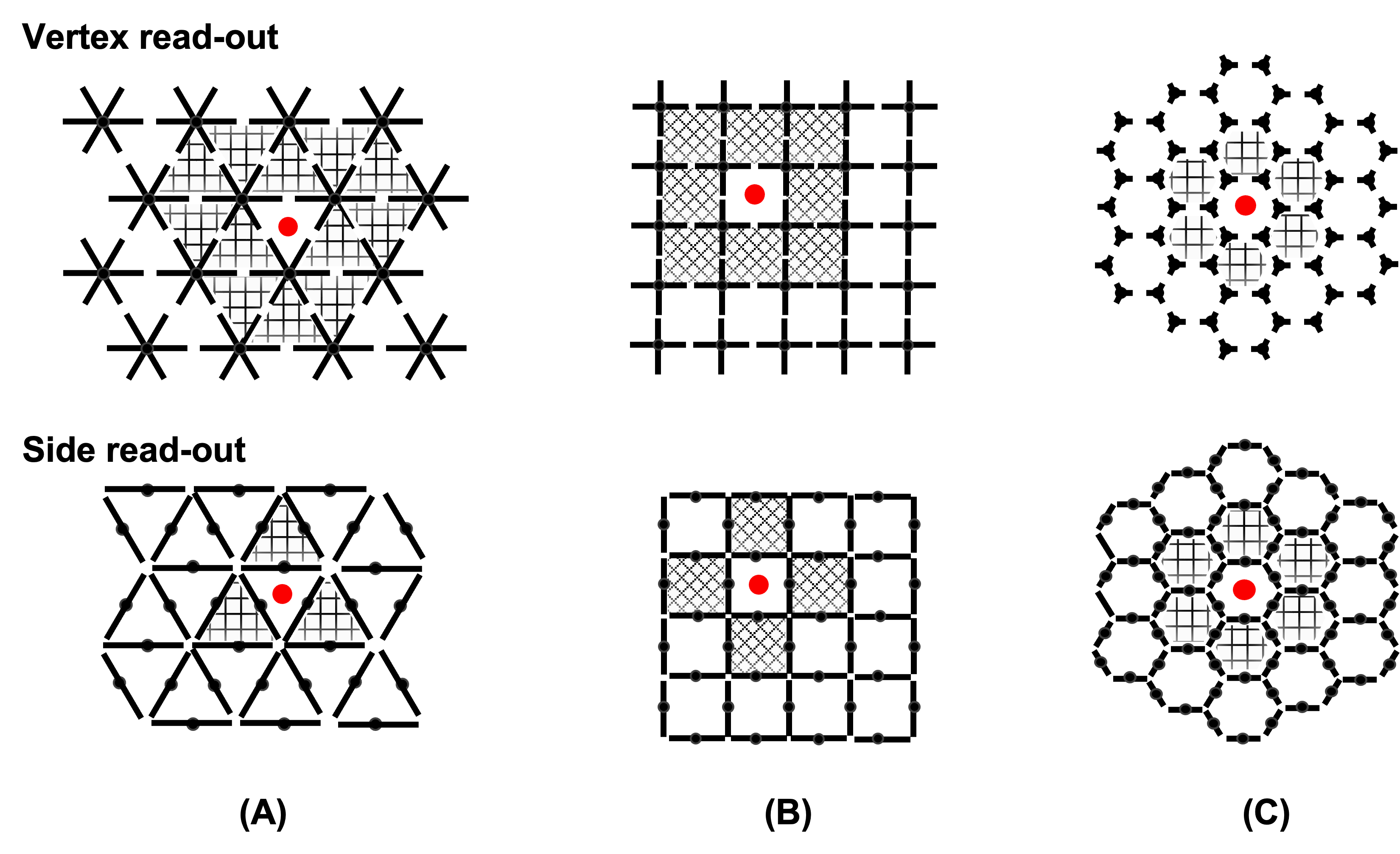}
\caption{Possible pixel shapes (triangle, square, and hexagon) and read-out electrode layouts (at the vertexes or at the sides). The most accurate resolution is obtained when the area around the hit  (shown as checkered) is not hit by a second particle.}
\label{fig:occupancy}
\end{center}
\end{figure}

In this section, a comparison among possible alternative pixel shapes (triangular, square, and hexagonal) and read-out electrode layouts (at the vertexes or at the sides) is presented. These layouts are shown in Figure~\ref{fig:occupancy}.

The most accurate resolution is obtained when the pixels around the impact point (red dot) are not hit by additional particles. For this reason, the determination of the optimum pixel size when using  RSD sensors should not be based solely on the spatial resolution but also on the sensor occupancy.   As a general rule,  power consumption (i.e. the number of read-out amplifiers) is minimized by using vertex electrodes. However, this choice maximizes the area that needs to be without additional particles.

\section{Conclusions}

This paper presents a detailed evaluation of the space and time resolutions of 50 \microns thick RSD sensors with cross-shaped electrodes, manufactured at FBK as part of the RSD2 production.
The studies, performed using a laser TCT setup, allow to estimate the performance of the sensors with charged particles, demonstrating the concurrent excellent space and time resolution over a large range of pixel sizes, from 200 \microns to 1300 \micron.

At gain = 30, the time resolution for all structures is between 35 - 40 ps, dominated by the Landau noise term, while the space resolution is about 3\% of the pitch size, dominated by the jitter term.
For equal spatial resolution, the RSD design reduces the number of read-out channels by about 50-100 with respect to sensors employing single-pixel read-out: this is a crucial feature to limit power consumption and to provide more space to fit the electronic circuits. 

This analysis also demonstrates that the $n^+$ resistive sheet  non-uniformity of the RSD sensors has a limited impact on the performance. 

\section*{Acknowledgments}

We thank our collaborators within RD50, ATLAS, and CMS, who participated in the development of UFSD. We kindly acknowledge the following funding agencies and collaborations: INFN-FBK agreement on sensor production; Dipartimenti di Eccellenza, Univ. of Torino (ex L. 232/2016, art. 1, cc. 314, 337), Italia; Ministero della Ricerca, Italia, PRIN 2017, Grant 2017L2XKTJ – 4DinSiDe; Ministero della Ricerca, Italia, FARE,  Grant R165xr8frt\_fare; Compagnia di San Paolo, Italia, Grant TRAPEZIO 2021; United States Department of Energy, USA, Grant DE-SC0010107.

\bibliography{RSD2_v5}

\end{document}